\documentclass[aps,twocolumn,groupedaddress,superscriptaddress,floatfix,pra]{revtex4-1}
\usepackage{amsmath}
\usepackage{mathtools}
\usepackage{amssymb}
\usepackage{graphicx}
\usepackage{textcomp}
\usepackage{bm}	
\usepackage{soul}
\usepackage{cancel}
\usepackage{ulem}
\usepackage{subcaption}

\usepackage[usenames,dvipsnames]{color}

\usepackage[braket, qm]{qcircuit}

\begin{document}

\title{Thermodynamic analysis of autonomous quantum systems}

\author{Tiago F. F. Santos}\email{t.santos0990@gmail.com}
\affiliation{Université Bourgogne Europe, CNRS, Laboratoire Interdisciplinaire Carnot de Bourgogne ICB UMR 6303, F-21000 Dijon, France}

\author{Camille Latune}
\affiliation{Université Bourgogne Europe, CNRS, Laboratoire Interdisciplinaire Carnot de Bourgogne ICB UMR 6303, F-21000 Dijon, France}

\date{\today}
\begin{abstract}

Traditional quantum thermodynamic frameworks associate work to energy exchanges induced by unitary transformations generated by external controls, and heat to energy exchanges induced by bath interaction. Recently, a framework was introduced aiming at extending the thermodynamic formalism to genuine quantum settings, also referred to as autonomous quantum systems: free from external controls, only quantum systems interacting with each other. In this paper, we apply such a thermodynamic framework to common experimental situations of interacting quantum systems. In situations where traditional frameworks detect only heat exchanges, the recent autonomous thermodynamic framework points at work exchanges based on two mechanisms: population inversion and coherence generation / consumption. Such mechanisms are well known in the literature for being related to work expenditure and extraction, in particular in relation with ergotropy, which emphasizes the relevance of the autonomous framework and the limitations of traditional ones. Furthermore, the autonomous framework also identifies a genuine non-unitary mechanism of work exchange related to athermality.  
Finally, in the semi-classical limit, the autonomous framework identifies all energy exchanges as pure work, but distinguishes between local work and interaction energy.
 Our results show that the autonomous framework provides a refined analysis of work exchange mechanisms in the quantum realm and serves as a consistent approach to analyze thermodynamic processes in realistic quantum devices.
\end{abstract}

\maketitle

\section{Introduction}
Quantum thermodynamics is a research field that combines quantum mechanics, thermodynamics, and quantum information theory to extend the concepts of classical thermodynamics into the quantum domain. Beyond its theoretical interest in understanding the mechanisms of energy exchange at the quantum level, where phenomena such as superposition, entanglement, and quantum fluctuations play a fundamental role, it also has direct implications for the development of quantum technologies~\cite{SingleIon2012, Pekola2015, BinderBook, GooldReview, Hofer2017, Xie2020, GLandi, AD, NonMarkovian2022, SzilardFT2024, Zambon, HeatWitness}, as well as for analyzing and improving the energetic footprint of quantm tecnhologies~\cite{A1, A2, A3, A4, A5}.

Despite significant progress over the past decade, quantum thermodynamics still lacks a universal formulation. Various theoretical frameworks based on thermodynamic resource theory, quantum information theory, and open quantum systems have been proposed. However, these approaches often yield conflicting results or interpretations in certain scenarios, particularly in non-equilibrium or strong-coupling regimes~\cite{1, esp, Spohn1978, Alicki1979, colla, maffei, dann, JM, PS}. In many cases, such discrepancies arise from fundamental differences in the assumptions underlying each framework. Moreover, some of these approaches are designed for specific classes of problems or for particular system configurations.

In Ref.~\cite{1} the authors presented a formalism to extend the laws of thermodynamics to autonomous quantum systems --systems that are not under the action of an external agent. This approach go beyond traditional definitions of pure heat and pure work sources commonly used in other formalisms, treating each subsystems on an equal footing as a hybrid source of both heat and work. In this scenario, they introduced consistent definitions of heat and work along with an expression for the second law of thermodynamics, all valid for arbitrary coupling strength and arbitrary number of interacting quantum systems.

In this work, we apply the formalism developed in \cite{1} to common experimental situations where traditional frameworks are not applicable or present some limitations: a two-level atom interacting with a single bosonic mode, a system of two interacting qubits, and a qubit interacting with a thermal bath.

 In the semiclassical limit, we show that this formalism reproduces the standard definitions used in quantum thermodynamics \cite{Spohn1978, Alicki1979}  and is consistent with other recent and established frameworks \cite{maffei, dann}, while distinguishing between local work and energy interaction. In the non-unitary regime, where quantum correlations play an important role, and therefore referred to as quantum regime in the following, we show that two unitary mechanisms of work exchange emerge: one related to population inversion, and one related to generation / consumption of coherences. Finally, we also point at a non-unitary work exchange mechanism in which the dimension of the quantum systems seems to be a determinant resource. 

This paper is organized as follows. In Section~II, we provide a brief review of the theoretical framework developed in Ref.~\cite{1}. In Section~III, we present a series of examples with both analytical and numerical solutions, considering semiclassical and quantum regimes, in order to illustrate some of the key features and implications of this formalism. Finally, in Section~IV, we summarize our main conclusions and discuss the potential applications and broader relevance of the framework introduced in Ref.~\cite{1}.

\section{Theoretical framework}\label{TF}

Before presenting our results, we will provide a brief review of the theoretical framework introduced in Ref.~\cite{1}. For this purpose, Consider two arbitrary quantum systems, \( A \) and \( B \), which interact with each other, with system \( A \) possibly being externally driven. The total Hamiltonian of the composite system, \( H(t) \), is given by $H(t) = H_A(t) + V_{AB}(t) + H_B$, where \( V_{AB}(t) \) represents the interaction between \( A \) and \( B \), and \( H_A(t) \) includes the external driving acting on system \( A \).

Assume that, at $t = 0$, the systems A and B are in a separable state $\rho(0) = \rho_A(t) \otimes\rho_B^{th}$, where $\rho_B^{th} = e^{-\beta_BH_B}/Z_B$ is the thermal equilibrium state with an inverse temperature $\beta_B$ and $Z_B = \operatorname{Tr} e^{-\beta_BH_B}$ being the partition function.

Under these assumptions, the authors in Ref.~\cite{esp} derived the following expression for the entropy production of system A at any time $t \geq 0$
\begin{align}\label{Esp}
    \sigma_A &= \Delta S_A + \beta_B \Delta U_B \nonumber \\
    &=I_{AB} + D[\rho_B(t)||\rho_B^{th}] \geq 0,
\end{align}
where $S_A$ is the Von Neumann entropy of system A, $S_{A(B)} = - \operatorname{Tr}[\rho_{A(B)}(t)\ln\rho_{A(B)}(t)]$, $U_B$ is the internal energy of system B, $U_{B(A)} = \operatorname{Tr}[\rho_{B(A)}(t)H_{B(A)}(t)]$, $I_{AB} = D[\rho(t)||\rho_A(t)\otimes\rho_B(t)]$ is the mutual information between A and B, built up during their joint evolution ($t > 0$), and $D[\rho_1(t)||\rho_2(t)] = \operatorname{Tr}\{\rho_1(t)[\ln\rho_1(t) - \ln\rho_2(t)]\}$ is the relative quantum entropy between the states $\rho_1(t)$ and $\rho_2(t)$.

Equation~\eqref{Esp} is always positive because both \( I_{AB} \) and \( D[\rho_B(t) \Vert \rho_B^{\text{th}}] \) are non-negative quantities. For this reason, Eq.~\eqref{Esp} is regarded as the second law of thermodynamics for open quantum systems. We emphasize this result here because it is directly connected to the central result of Ref.~\cite{1}, which we will discuss shortly.

Although the expressions we will show in the following hold for a time-dependent Hamiltonian of the form $H(t) = H_A(t) + V_{AB}(t) + H_B$, from now on we will assume that the global Hamiltonian, $H(t)$, is time-independent. This assumption underlies all subsequent discussions and results, as our focus is on autonomous quantum systems—that is, systems that evolve without the action of an external agent.

Now, let us recall the definitions of heat and work introduced in \cite{1}. Consider that quantum systems A and B are initially in an arbitrary uncorrelated state $\rho(0) = \rho_A(0) \otimes \rho_B(0)$, where each of them can act as a source of both heat and work. The Hamiltonian of the global system is given by $H = H_A + V_{AB} + H_B$, where $V_{AB}$ represents the interaction between A and B. To split the energy exchange between systems A and B into heat and work contributions, the authors introduced the quantity~\cite{1, TE1, TE2}
\begin{equation}\label{eth}
U^{th}_j(t) = \operatorname{Tr}[\rho_j^{th}(t)H_j], \hskip 0.5cm j = A, B
\end{equation}
which is referred to as the thermal energy of system $j$, where $\rho_j^{th}(t) = e^{-\beta_j(t)H_j}/Z_j$ denotes the thermal state of system $j$ constructed to have the same von Neumann entropy as the actual state $\rho_j(t)$, i.e.,
\begin{equation}\label{ent}
S[\rho_j^{th}(t)] = S[\rho_j(t)].
\end{equation}
$\beta_j(t)$ is the effective inverse temperature defined by Eq. \eqref{ent}. Note that Eq. \eqref{ent} always admits a unique solution, thereby uniquely defining the thermal state $\rho_j^{th}(t)$ (except for  systems  with degenerate ground state, see ~\cite{comment}). By defining heat as
\begin{equation}\label{heat}
Q_j = - \Delta U^{th}j(t),
\end{equation}
and using Eq.~\eqref{Esp}, the authors in \cite{1} obtained their central result:
\begin{align}
\sigma_j &= \Delta S_j - \beta_{i\neq j}(0) Q_{i\neq j} \nonumber \\
&=I_{AB}(t) + D[\rho_j^{th}(t)||\rho_j^{th}(0)] \geq 0 \label{CLL}
\end{align}
It is worth noting that the main difference between Eqs.~\eqref{Esp} and~\eqref{CLL} is that, in Eq.~\eqref{CLL}, B is allowed to be initially in an arbitrary state. This implies that both systems can simultaneously act as sources of heat and work.

In this scenario, Eq.~\eqref{CLL} represents the second law of thermodynamics for the quantum systems A and B. Since it depends on the initial effective inverse temperature $\beta_j(0)$, certain constraints are imposed on the heat flow through Eq.~\eqref{CLL}. It is also worth noting that, because heat is defined as the variation of the thermal energy, it is directly related to the entropy change. In fact, it is straightforward to show that
\begin{equation}\label{2law}
\dot{S}_j = - \beta_j(t)\dot{Q}_j.
\end{equation}

We can define work from the first law of thermodynamics as
\begin{equation}\label{work}
W_j(t) = -\Delta U_j - Q_j(t),
\end{equation}
where $U_j = \operatorname{Tr}[\rho_j(t)H_j]$ is the average energy of system $j$. Here, it is worth mentioning that due to the choice of sign convention, $Q > 0 \hskip 0.05cm (Q < 0)$ corresponds to the heat provided by (absorbed by) the system, and $W > 0 \hskip 0.05cm (W < 0)$ corresponds to the work provided (received) by the system. We can rewrite Eq.~\eqref{work} as
\begin{equation}\label{work2}
W_j(t) = - \operatorname{Tr}{[\rho_j(t) - \rho_j^{th}(t)]H_j} + \operatorname{Tr}{[\rho_j(0) - \rho_j^{th}(0)]H_j},
\end{equation}
and define $\Xi_j(t) := \operatorname{Tr}\{[\rho_j(t) - \rho_j^{th}(t)]H_j\}$. Then,
\begin{equation}\label{exergy}
W_j(t) = - \Xi_j(t) + \Xi_j(0) = - \Delta \Xi.
\end{equation}

Now, let us define a passive state $\pi_j(t) = \sum_i r_i(t)\ket{E_i}\bra{E_i}$ with the same spectrum as $\rho_j(t)$, where ${\ket{E_i}}$ is the energy eigenbasis of system $j$. Note that, since it has the same spectrum as $\rho_j(t)$, $\pi_j(t)$ also has the same entropy, i.e., $S[\rho_j(t)] = S[\pi_j(t)]$. We can use the internal energy associated with this passive state, $U_P(t) = \operatorname{Tr}[H_j\pi_j(t)]$, to rewrite $\Xi_j(t)$ as (see Appendix \ref{AppA1} for details)
\begin{align}\label{ex}
\Xi_j(t) &=\operatorname{Tr}\{[\rho_j(t) - \pi(t)]H_j\} + \operatorname{Tr}\{[\pi(t) - \rho_j^{th}(t)]H_j\} \nonumber \\
&= \mathcal{E}_j(t) + \frac{1}{\beta_j(t)} D[\pi_j(t)||\rho_j^{th}(t)] \geq 0,
\end{align}
where the first term on the right-hand side of Eq. \eqref{ex} is the ergotropy, $\mathcal{E}_j(t)$, which corresponds to the maximum work that can be extracted unitarily~\cite{Ergotropy}.

Note that, Eq. \eqref{ex} is always positive because both the ergotropy and the relative quantum entropy are positive quantities. This expression also shows that $\Xi(t)$ is always greater than the ergotropy. We call $\Xi(t)$ exergy~\cite{exergy1, exergy2}, and one can show (see~\cite{1}) that it corresponds to the energy that can be extracted with a thermal bath at inverse temperature $\beta_j(t)$. Substituting Eq.~\eqref{ex} into Eq.~\eqref{work2}, we obtain
\begin{equation}\label{exergy3}
W_j = - \Delta \mathcal{E}_j(t) - \Delta\left\{\frac{1}{\beta_j(t)} D[\pi_j(t)||\rho_j^{th}(t)]\right\}.
\end{equation}
Crucially, Eq. \eqref{exergy3} highlights that work can be produced through two different mechanisms: unitary mechanisms, associated with the ergotropy, and non-unitary mechanisms, associated with the relative entropy. Thus, as long as the system is in a passive state different from the thermal state, it can still provide work, as also pointed out in~\cite{exergy2}.

\section{Results}

In this section, we present some results and aspects of the above framework for both the semiclassical and quantum regimes. In the semi-classical regime, we compare the formalism described in the previous section with what we call the standard framework, and we show that there are differences between them. We also show that it is possible to connect the framework developed in \cite{1} with other recent approaches in the semi-classical regime. In the quantum regime, we use specific examples to highlight emerging distinctive features.

\subsection{Semi-classical regime}

Consider that two systems, A and B, are interacting and isolated from any other external systems. The full Hamiltonian of the global system is given by $H = H_A + V_{AB} + H_B = H_A + g\bar{V}_{AB} + H_B$, where $V_{AB} = g\bar{V}_{AB}$ represents the interaction between A and B and $g$ is the coupling strength. Now, consider that the correlations generated by the interaction term $V_{AB}$ can be neglected, so the joint state of the system is always a separable state, $\rho(t) \approx \rho_A(t) \otimes \rho_B(t)$. This is what we refer to as the semiclassical limit (or regime). This can happen, for instance, for timescales much shorter than $g^{-1}$, or when either B or A becomes a "classical system", for example, a harmonic oscillator in a large coherent state in the weak coupling limit \cite{QO1, QO2,Jha}. In this scenario, the local dynamics becomes
\begin{align}
    \dot{\rho}_A(t) = - \frac{i}{\hbar}[H_A^{eff}(t), \rho_A(t)] \label{dyna}, \\
    \dot{\rho}_B(t) = - \frac{i}{\hbar}[H_B^{eff}(t), \rho_B(t)],
\end{align}
where $H_A^{eff}(t)$ and $H_B^{eff}(t)$ are effective Hamiltonians
\begin{align}\label{heff}
    H_A^{eff}(t) = H_A + \operatorname{Tr}_B[\rho_B(t)V_{AB}] \\
    H_B^{eff}(t) = H_B + \operatorname{Tr}_A[\rho_A(t)V_{AB}].
\end{align}
From now on, we will consider that system B plays the role of a classical system and that system A is the system of interest.

From Eq. \eqref{heat}, we see that there is no heat exchange, $Q_A(t) = 0$, since $S_A$ is constant in time. Then, from Eq. \eqref{work}, we get
\begin{equation}\label{wuni}
    W_A (t) = - \Delta U_A(t) = -\operatorname{Tr}\{H_A[\rho(t) - \rho(0)]\}.
\end{equation}
Note that in the semi-classical regime, A is seen to be driven by the semiclassical system B, and the local Hamiltonian of A is the driven Hamiltonian $H_A^{eff}(t)$. Applying the standard definitions~\cite{Spohn1978, Alicki1979} to the effective dynamics Eq. (12), one obtains,
\begin{align}\label{wst}
    \dot{W}_A^{st} &= \operatorname{Tr}\left[\rho_A(t)\dot{H}_A^{eff}(t)\right] \\
    \dot{Q}_A^{st} &= \operatorname{Tr}\left[\dot{\rho}_A(t)H_A^{eff}(t)\right] \label{qst} \\
    U^{st}_A(t) &= \operatorname{Tr}\left[\rho_A(t)H_A^{eff}(t)\right] = Q_A^{st} + W_A^{st}.
\end{align}
Note that in the semiclassical regime there is no heat exchange,$\dot{Q}_A^{st} = 0$. This follows directly by combining Eqs. \eqref{dyna} and \eqref{qst}.

Now, Using Eq. \eqref{heff} to calculate the internal energy of $A$, we obtain
\begin{equation}
    U^{st}(t) = U_A + E_{int},
\end{equation}
where $E_{int} = \operatorname{Tr}\left[\rho(t)V_{AB}\right]$ is the interaction energy. Then,
\begin{equation}\label{weint}
    \dot{W}^{st} = \dot{W}_A + \dot{E}_{int}.
\end{equation}
This means that if the interaction energy is constant, $\dot{E}_{int} = 0$, the standard definition of work coincides with $\dot W_A$, the definition from the autonomous formalism \cite{1}. A simple example illustrating this is the Jaynes-Cummings model, where system B is a harmonic oscillator in a large coherent state and system A is a two-level system. In this case, if the systems are resonant $\dot{E}_{int} = 0$, then $\dot{W}^{st} = \dot{W}_A \Rightarrow W^{st} = W_A$.

It is worth mentioning that Eq. \eqref{weint} reinforces the idea that the interaction energy plays the role of a (non-local) source of work. It goes along with the observation that the switch on and off of the interaction has a cost equal to the variation of the interaction energy, $W_{on} +W_{off} = \Delta E_{int}$. In this scenario, we see that the full quantum framework developed in Ref.~\cite{1} seems to be more precise, as it distinguishes between pure local work and the contribution arising from the interaction energy.

The framework shown here can also be connected to other recent frameworks in the semiclassical regime. For example, in the framework "MCA" (named after the names of the authors)~\cite{maffei}, the authors considered two interacting quantum systems, A and B, and split their energy change, $\dot{U}_j(t)$, into two contributions, which they refer to as work flow and coherent energy flux.

The energy flow for the system $j$ is given by
\begin{align}\label{um}
    \dot{U}_j(t) &= - i\hskip 0.02cm \operatorname{Tr}_j \left\{[H_j, \mathcal{H}_j(t)]\rho_j(t)\right\} \nonumber \\
    &- i\hskip 0.02cm \operatorname{Tr}_{l\neq j}\left\{H_j\left[V_{AB}(t), \chi(t)\right]\right\},
\end{align}
where, as previously, $U_j(t) = \operatorname{Tr}[H_j\rho_j(t)]$, $H_j$ is the free Hamiltonian of system $j$, $\mathcal{H}_j(t) = \operatorname{Tr}_{l \neq j}[V_{AB}(t)\rho_l(t)]$ is an effective driving term, $\rho_j(t)$ is the density matrix of system $j = A,B$, $V_{AB}(t)$ is the interaction Hamiltonian between A and B, and  $\chi(t) = \rho(t) - \rho_A(t) \otimes \rho_B(t)$ is the correlation matrix ($\rho(t)$ is the joint state of AB). From Eq. \eqref{um} they defined the work rate as
\begin{equation}\label{wm}
    \dot{W}_j^{MCA}(t) = - i\hskip 0.02cm \operatorname{Tr}_j \left\{[H_j, \mathcal{H}_j(t)]\rho_j(t)\right\}
\end{equation}
and the correlation energy flow as
\begin{equation}
    \dot{Q}_j^{MCA}(t) = - i\hskip 0.02cm \operatorname{Tr}_{l\neq j}\left\{H_j\left[V_{AB}(t), \chi(t)\right]\right\}.
\end{equation}
In the semiclassical regime $\rho(t) \approx \rho^A(t) \otimes \rho^B(t)$, implying $\chi(t) = 0$, $\dot{Q}^j = 0$, and a dynamics given by Eq.~\eqref{dyna}. Then, using Eqs. \eqref{wm} and \eqref{dyna}, we obtain
\begin{equation}\label{wmm}
    \dot{W}_j^{MCA} = \operatorname{Tr}\left[H_A\dot{\rho}_A(t)\right] = \dot{U}^j(t),
\end{equation}
as expected for the semiclassical regime. Comparing the expression above with the standard definition of work, Eq. \eqref{wst}, we get
\begin{equation}\label{weint2}
    \dot{W}^{st} = \dot{W}_j^{MCA} + \dot{E}_{int},
\end{equation}
implying that the definitions of work in~\cite{1} and in~\cite{maffei} coincide in the semiclassical regime.

Another interesting case is the framework developed in~\cite{dann}. In this work, the authors defined heat and work for autonomous systems and established a connection between their definitions and the standard definitions of work and heat (For details see section IV - VII in~\cite{dann}). They obtained a general expression for the production of work in the autonomous case, $\dot{W}_a^A$, in terms of the standard definition of the production of work, $\dot{W}^{st}$,
\begin{equation}
    \dot{W}_a^A = \dot{W}^{st} + \mathcal{F}(t),
\end{equation}
where $\mathcal{F}(t)$ is a function that depends on the correlation between A and B. In the semi-classical regime, we have that $\mathcal{F}(t) = 0$. Then, the autonomous work production definition is therefore different from the work production in the unitary regime in Eqs. \eqref{weint} and \eqref{weint2}. This means that, within their definition, it is not possible to distinguish between the purely local work contribution and the interaction energy in the unitary regime, unlike the frameworks in Refs.~\cite{1, maffei}.

In the next section, we investigate the quantum regime. To this end, we consider examples from experimental situations like a two-level atom interacting with electromagnetic mode, qubit–qubit interaction and spontaneous emission phenomenon.

\subsection{Non-unitary regime}

\subsubsection{Qubit-Harmonic oscillator interaction}

Consider a typical experimental situation ~\cite{QO1, C1, C2} of a two-level system A, interacting with an electromagnetic mode represented by a quantum harmonic oscillator (QHO), system B, via a Jaynes cummings model. The full Hamiltonian of the composite system is given by
\begin{equation}
    H = \frac{\hbar \omega_A}{2} \sigma_z + \hbar \omega_B a^{\dagger}a + g\,(\sigma_+a + \sigma_-a^{\dagger}),
\end{equation}
where $\sigma_z$ is the Pauli matrix, the qubit has energy levels $\ket{g}$ and $\ket{e}$, with $E_e > E_g =0$, $\sigma_- = \sigma_+^{\dagger} = \ket{g}\bra{e}$ and $a^{\dagger}$ ($a$) is the usual creation (annihilation) operator. The subsystems are assumed to be resonant ($\omega_A = \omega_B$). Here, we will analyze the thermodynamics of the system in four different cases: in the first case, we will consider the qubit initially in the ground state, $\ket{g}$, and the QHO initially in a coherent state with a large number of photons ($\bar{n} \gg 1$). In all subsequent cases, the QHO is assumed to be initially in the vacuum state, $\ket{0}$, while the qubit, in the second case, will initially be in the excited state, in the third case, it will initially be in a superposition state, $\ket{\psi_A(0)} = \frac{\ket{g} + \ket{e}}{\sqrt{2}}$, and finally in the fourth case, the qubit will initially be in a Gibbs state.

\begin{figure}[h!]
    \begin{subfigure}[b]{0.8\linewidth}
        \includegraphics[width=\linewidth]{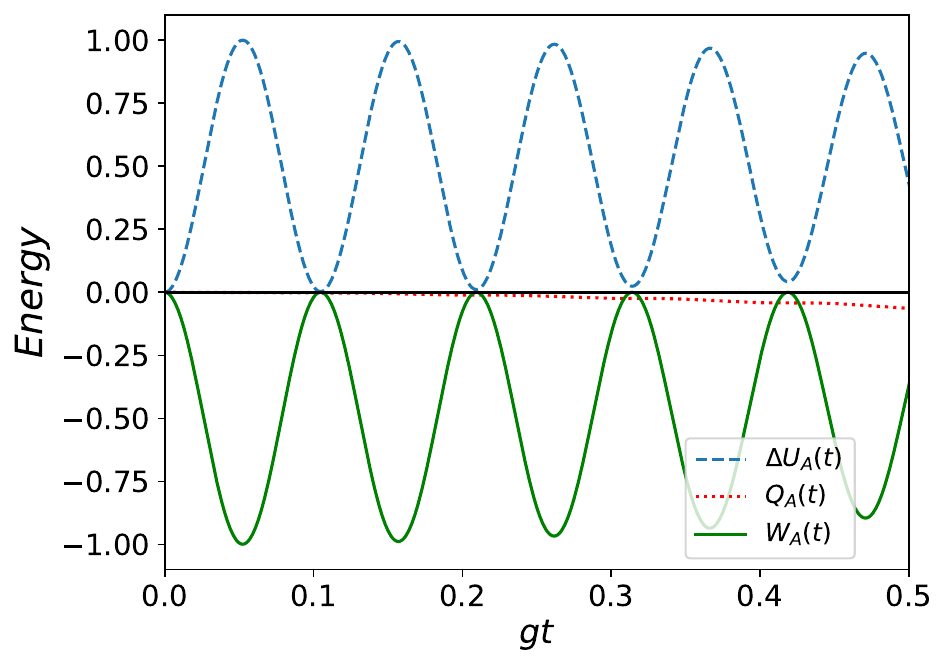}
        \caption{Qubit (System A).}
        \label{fig:sub01}
    \end{subfigure}
    
    \vspace{0.5cm} 

    \begin{subfigure}[b]{0.8\linewidth}
        \includegraphics[width=\linewidth]{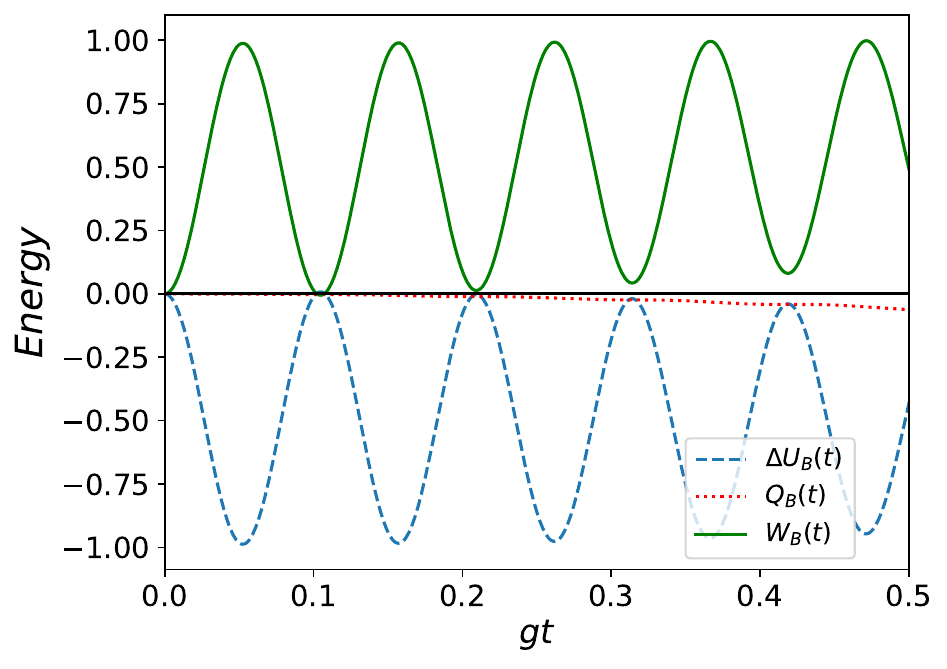}
        \caption{Quantum Harmonic Oscillator (System B).}
        \label{fig:sub02}
    \end{subfigure}
    \caption{Plot of the internal energy, heat, and work for the qubit (A) and the QHO (B) as functions of the dimensionless time $gt$. All thermodynamic quantities are expressed in units of $\hbar \omega$; this convention is used throughout all subsequent figures. The qubit is initially in the ground state and QMO is initially in a coherent state $\ket{\psi_{HO}}(0) = \ket{\alpha}$. Parameters: $\omega_A = \omega_B$, $\alpha = 30$ and $g = 0.01 \omega_A$.}
    \label{fig0}
\end{figure}

In Fig. \ref{fig0}, we present the thermodynamic quantities corresponding to the first case described above. In this scenario, the QHO is initialized in the coherent state $\ket{\psi_{HO}}(0) = \ket{\alpha}$, with $\bar{n} = |\alpha|^2 = 900 \hskip 0.1cm (\alpha = 30)$ reproducing the experimental situation of a two-level system coherently driven by a laser, used on daily basis in many experimental platforms to control and manipulate qubits~\cite{QO1, C1, C2}.

From Figs. \ref{fig:sub01} and \ref{fig:sub02}, we see that initially, in the dimensionless time interval between $gt = 0$ and $gt = 0.1$, approximately, there is no heat exchange between the systems. This occurs because, within this time scale, the QHO behaves like a classical system, so the dynamics of the qubit is governed by Eq. \eqref{dyna}. This corresponds to the well-known semiclassical regime used experimentally for coherent drive. However, for $gt > 0.1$, we start to see heat exchange between the systems, which means that the semiclassical approximation starts to break down, and the dynamics of the qubit is no longer unitary, as also pointed out in \cite{Jha}.  Interestingly, in both the qubit and the QHO, the heat flow is not oscillating as does the work flow, but instead increases slowly and steadily, as dissipation slowly comes in. Note also that we observe a damping effect in the work produced for $gt>0.1$.

In conclusion, the thermodynamic analysis reproduces what expected from daily experience in many experimental platforms: the qubit receives work from the laser. However, after many Rabi oscillations, when $gt$ is not much smaller that 1, the break down of the coherent drive window reflects on the work transfer which starts to be degraded together with the emergence of heat flow.

In the following, we analyze typical situations deep into the quantum regime,  where the QHO is initially in the vacuum state. Considering first that the qubit is initially in the excited state, the time evolution of the joint state is of the form
\begin{equation}\label{cdyn}
    \ket{\psi(t)} = c_1(t)\ket{e,0} + c_2(t) \ket{g,1},
\end{equation}
where $c_1(0) = 1$ and $c_2(0) = 0$. Tracing over the QHO, system B, we obtain for the reduced state of the qubit, system A,
\begin{equation}\label{A}
    \rho_A(t) = p_e(t) \ket{e}\bra{e} + p_g(t)\ket{g}\bra{g},
\end{equation}
$p_e(t) = |c_1(t)|^2 = \cos^2(gt)$ and $p_g(t) = |c_2(t)|^2 = \sin^2(gt)$ (See~\cite{QO1} for details). Similarly, tracing over the qubit yields the reduced state of the QHO,
\begin{equation}\label{B}
    \rho_B(t) = p_1(t) \ket{1}\bra{1} + p_0(t)\ket{0}\bra{0},
\end{equation}
with $p_1(t) = p_g(t)$ and $p_0(t) = p_e(t)$.

\begin{figure}[h!]
    \centering
    \begin{subfigure}[b]{0.8\linewidth}
        \centering
        \includegraphics[width=\linewidth]{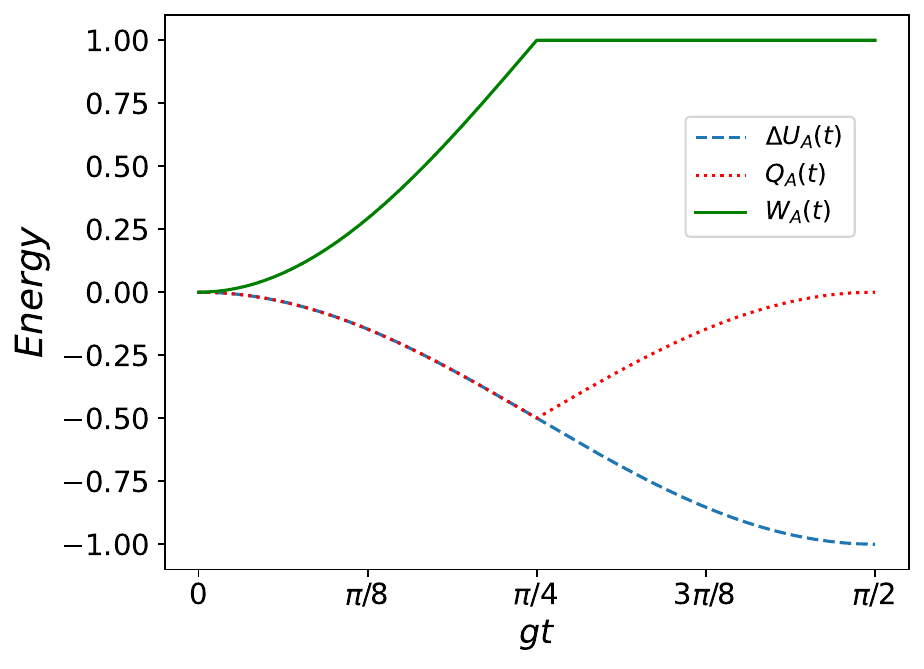}
        \caption{Qubit (System A).}
        \label{fig:sub31}
    \end{subfigure}
    
    \vspace{0.5cm} 

    \begin{subfigure}[b]{0.8\linewidth}
        \centering
        \includegraphics[width=\linewidth]{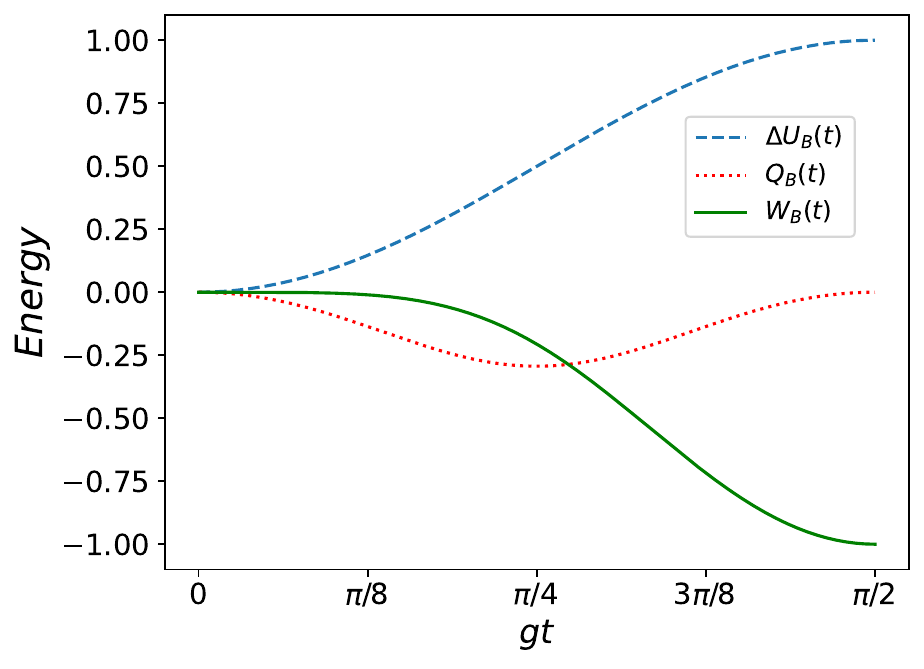}
        \caption{Quantum Harmonic Oscillator (System B).}
        \label{fig:sub32}
    \end{subfigure}
    \caption{ Plot of the thermodynamical quantities for both the qubit (A) and the QHO (B), as a function of time . Parameters: $\omega_A = \omega_B$ and $g = 0.01 \omega_A$.}
    \label{fig3}
\end{figure}

\begin{figure}[h!]
    \centering
    \begin{subfigure}[b]{0.8\linewidth}
        \centering
        \includegraphics[width=\linewidth]{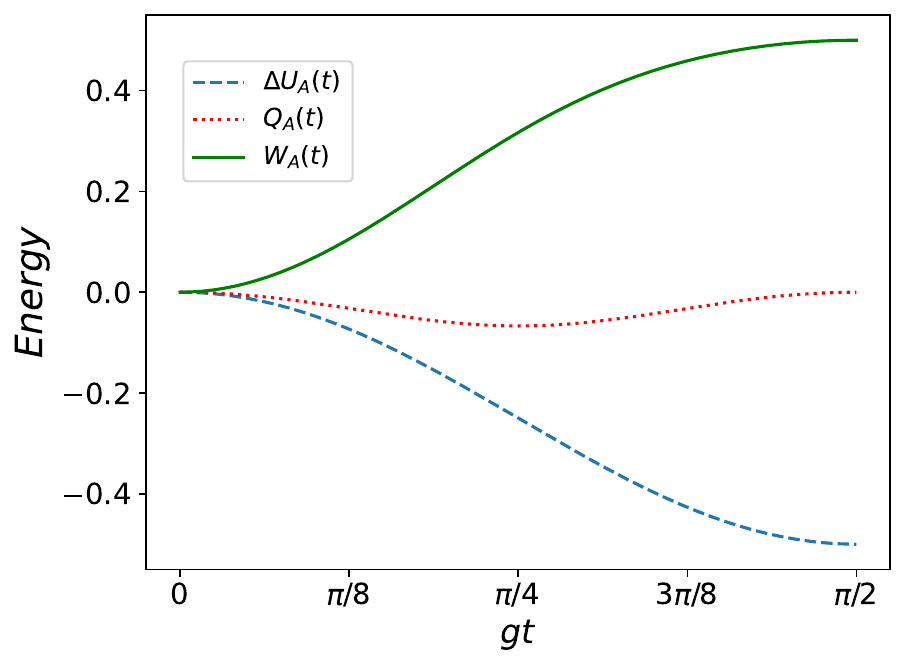}
        \caption{Qubit (System A).}
        \label{fig:sub41}
    \end{subfigure}
    
    \vspace{0.5cm} 

    \begin{subfigure}[b]{0.8\linewidth}
        \centering
        \includegraphics[width=\linewidth]{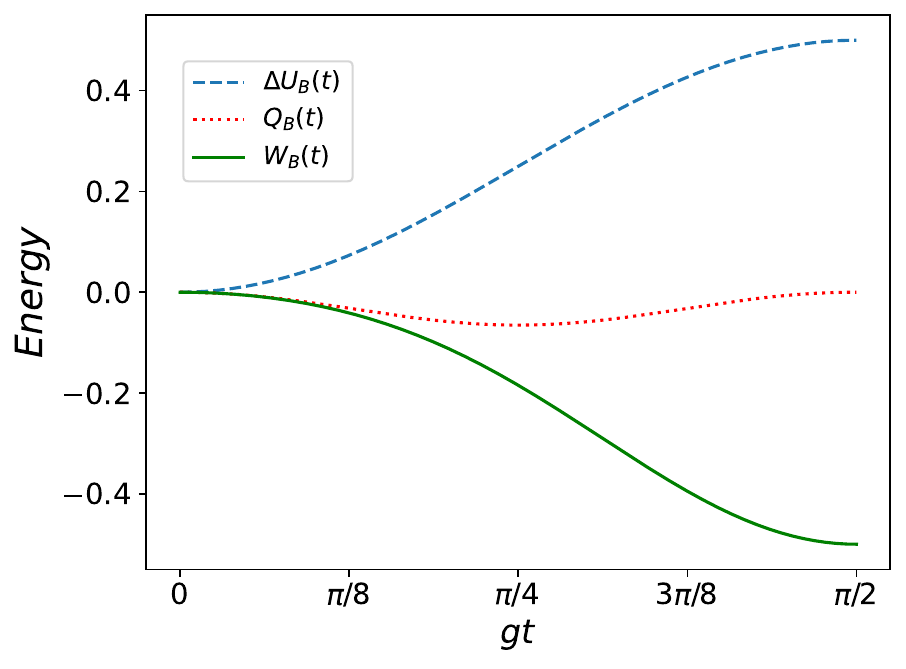}
        \caption{Quantum Harmonic Oscillator (System B).}
        \label{fig:sub42}
    \end{subfigure}
    \caption{ Plot of the internal energy, heat and work for the qubit and the QHO, as a function of time. The qubit is initially in the state $\ket{\psi_A(0)} = \frac{1}{\sqrt{2}}(\ket{e} + \ket{g})$. Parameters: $\omega_A = \omega_B$ and $g = 0.01 \omega_A$.}
    \label{fig4}
\end{figure}

\begin{figure}
    \centering
    \begin{subfigure}[b]{0.8\linewidth}
        \centering
        \includegraphics[width=\linewidth]{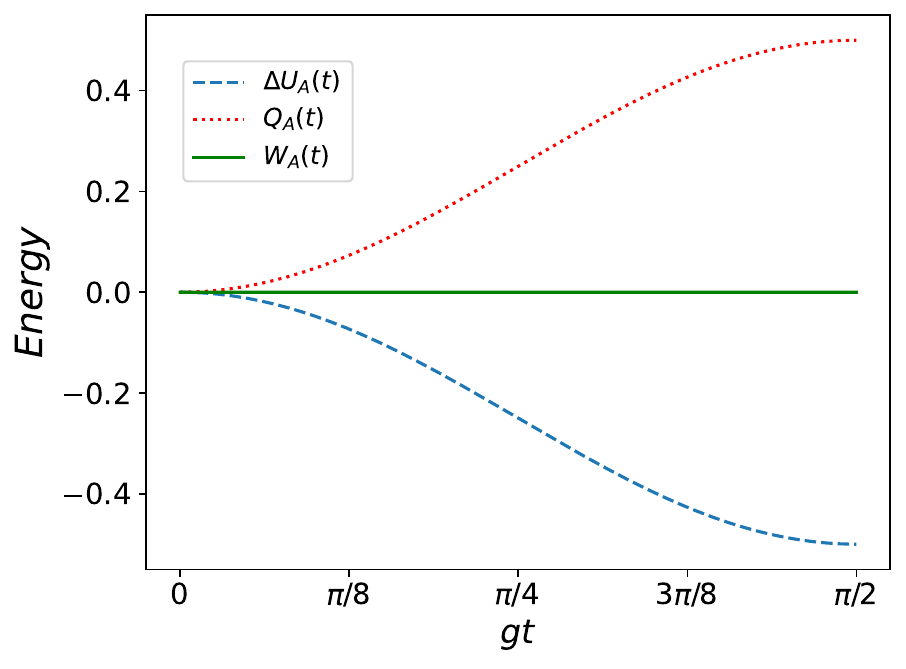}
        \caption{Qubit (System A).}
        \label{fig:sub51}
    \end{subfigure}
    
    \vspace{0.5cm} 

    \begin{subfigure}[b]{0.8\linewidth}
        \centering
        \includegraphics[width=\linewidth]{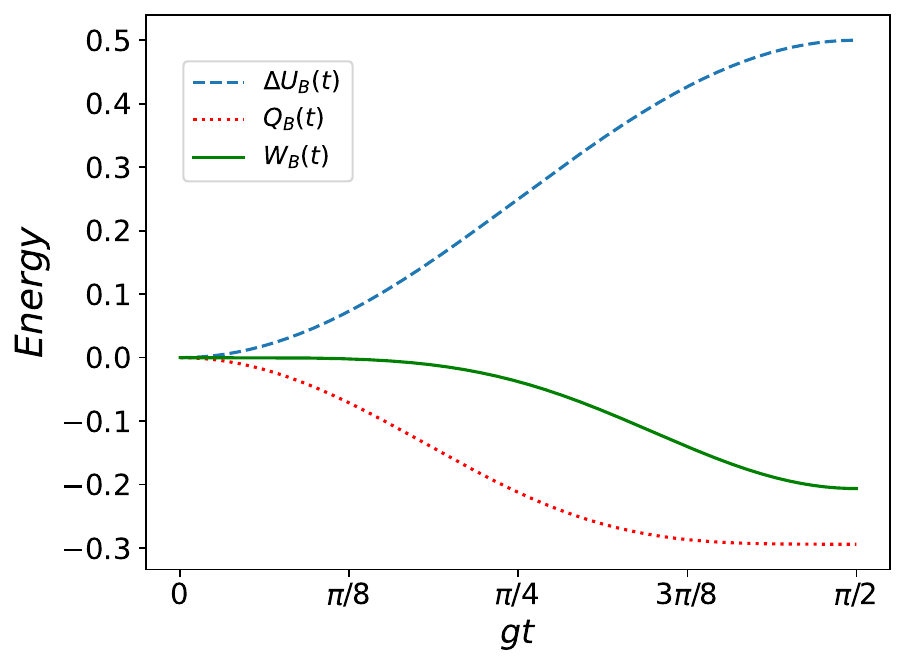}
        \caption{Quantum Harmonic Oscillator (System B).}
        \label{fig:sub52}
    \end{subfigure}
    \caption{Plot of the thermodynamical quantities for the qubit and the QHO, as a function of time. The qubit is initially in a thermal state $\rho_A(0) = \frac{1}{2}(\ket{g}\bra{g} + \ket{e}\bra{e})$. Parameters: $\omega_A = \omega_B$ and $g = 0.01 \omega_A$.}
    \label{fig5}
\end{figure}

As general observation, looking only at the energy change of the qubit (see Figs. \ref{fig3}, \ref{fig4}, and \ref{fig5}), one would see a behaviour similar to the previous situation since both energy curves correspond to squared sinusoidal functions (except that the timescale of the oscillations is not the same). However, the decomposition into heat and work exchanges is very different. This is due to the emergence of correlations. This also illustrates the well-known statement that heat and work depend strongly on the path. Additionally, the symmetry between the qubit and the QHO thermodynamic quantities observed in the previous situation Fig. \ref{fig0} is now broken, illustrating also the two above points. In the following, we provide a more detailed description of the plots of Figs. \ref{fig3}, \ref{fig4}, and \ref{fig5}.   \\
In Fig.~\eqref{fig3}, we present the thermodynamic quantities of interest for both the qubit and the QHO.  We restrict our analysis to the time interval \( 0 \leq t \leq \tau = \pi / 2g \), since the dynamics and consequently the energy exchange are periodic (See Appendix \ref{AppA2} for details).
In Fig.~\ref{fig:sub31}, we observe that during the interval \( 0 \leq t \leq \pi / 4g \), where \( p_e(t) \geq p_g(t) \), work  $W_A(t)$ is provided by the qubit while it absorbs heat, indicating that the purity of the qubit is degraded by the correlations build up with the QHO. 

For \( \pi / 4g < t \leq \pi / 2g \), the qubit ceases to produce work because \( p_g(t) \) becomes larger than \( p_e(t) \). During this time interval, the qubit is in a thermal state, and therefore no resources are available to produce work. Moreover, the system begins to release heat to the bath as it evolves from the maximally mixed state to the ground state \( \ket{g} \).

In Fig.~\eqref{fig:sub32}, the heat flow $Q_B$ mainly differs from the qubit one $Q_A$ around the maximal deep at $gt = \pi/4$. In particular, the absence of a sharp transition in $Q_B$ arises from the fact that no population-inversion mechanism is present. Regarding the work in the QHO, we see that even though \( p_0(t) \geq p_1(t) \) for \( t \leq \pi / 4g \), there is work being done on the QHO. This occurs because during this time interval, the system is in a passive state but not a thermal state, unlike the qubit. It is a direct consequence of the difference in the Hilbert space dimension of the two systems. Such work gain by the QHO is an illustration of the second term of Eq.~\eqref{ex}, meaning that work can be gained from purely non-unitary mechanisms.

As a third example, in order to investigate the effects of coherences in the systems, we consider that the qubit (system A) is initially prepared in a superposition state, $\ket{\psi_A(0)} = \frac{1}{\sqrt{2}}(\ket{e} + \ket{g})$, while the QHO is still initially in a vacuum state.

In Fig.~\ref{fig4}, we show the thermodynamic quantities for both systems. First of all, one can notice the symmetry / anti-symmetry between the qubit and QHO flows. Additionally, we can note that both systems absorb heat as they evolve from pure local states to mixed local states at \( t = \pi / 4g \), and subsequently, they release heat as they are purified, i.e., as they evolve toward local pure states at \( t = \pi / 2g \). Regarding work, we see that work is continuously extracted from the qubit throughout the entire evolution, reaching its maximum at \( t = \pi / 2g \), while work is continuously provided to the QHO, reaching its maximum value at the same time. This behavior arises because the coherence initially contained in the qubit is fully transferred to the QHO over the time interval \( 0 \leq t \leq \pi / 2g \).
This figure illustrates that coherence can indeed be used as a resource for work generation: generating coherence corresponds to providing work to the system, while consuming coherence enables the extraction of work from the system.

It is worth commenting that the main difference between the third and the second example lies in the work production mechanism. In the second example, population inversion and the dimension of the systems served as the mechanisms responsible for work generation, while in the third example coherence acts as the mechanism responsible for work generation.

As a fourth and final example, we consider the case where the qubit is initially in a maximally mixed state, i.e., a thermal state at infinite temperature, while the QHO is initially in the vacuum state. In Fig. \ref{fig5}, we observe that, in this scenario, the qubit only provides heat to the QHO, whereas the QHO not only absorbs heat but also receives work.
The mechanism responsible for this work production is the same as in the second example: the difference in the dimensionality of the two systems, illustrating again that work exchange is possible through purely non-unitary mechanisms, as indicated by the second term of Eq. \eqref{ex}.

Additionally, it is worth mentioning that if the QHO were initially in a thermal state with a nonzero temperature, some work production would still occur, but with a smaller magnitude since the total energy exchange in that scenario would be reduced. However, when both systems have the same Hilbert space dimension and both start in thermal states at different temperatures, only heat exchange is observed. This result re-enforces our previous observations and suggests that the dimensional difference between the Hilbert spaces of quantum systems may itself serve as a resource for work production, even when the systems are initially in thermal states. 

In the next example, we replace QHO by a qubit and analyze the thermodynamic behavior of the system over the same time interval as considered above.

\subsubsection{Qubit-Qubit interaction}

Now, let us consider two qubits, \(A\) and \(B\), each with energy levels \(\ket{g}\) and \(\ket{e}\), and corresponding energies \(E_e > E_g = 0\). The total Hamiltonian of the composite system is given by
\begin{equation}
    H = \hbar\omega_A \sigma_{ee}^{(A)} + \hbar\omega_B \sigma_{ee}^{(B)} + \hbar g\,(\sigma_+^{(A)}\sigma_-^{(B)} +  \sigma_-^{(A)}\sigma_+^{(B)}),
\end{equation}
where $\sigma_{ee} = \ket{e}\bra{e}$, $\sigma_- = \sigma_+^{\dagger} = \ket{g}\bra{e}$ and the qubits are assumed to be resonant ($\omega_A = \omega_B$).

In the first example, we consider that qubit A is initially in the excited state $\ket{e}$, while qubit B is initially in the ground state $\ket{g}$. In this scenario, the reduced state of each qubit is given by
\begin{equation}
    \rho_A(t) = p_e^A(t) \ket{e}\bra{e} + p_g^A(t)\ket{g}\bra{g}
\end{equation}
and
\begin{equation}
    \rho_B(t) = p_g^B(t) \ket{e}\bra{e} + p_e^B(t)\ket{g}\bra{g},
\end{equation}
where $p_e^A(t) = p_g^B(t)$ and $p_g^A(t) = p_e^B(t)$. Note that both qubits are always in a diagonal state, therefore if $p_e(t) \geq p_g(t)$ we have population inversion in the system.

\begin{figure}
    \centering
    \begin{subfigure}[b]{0.8\linewidth}
        \centering
        \includegraphics[width=\linewidth]{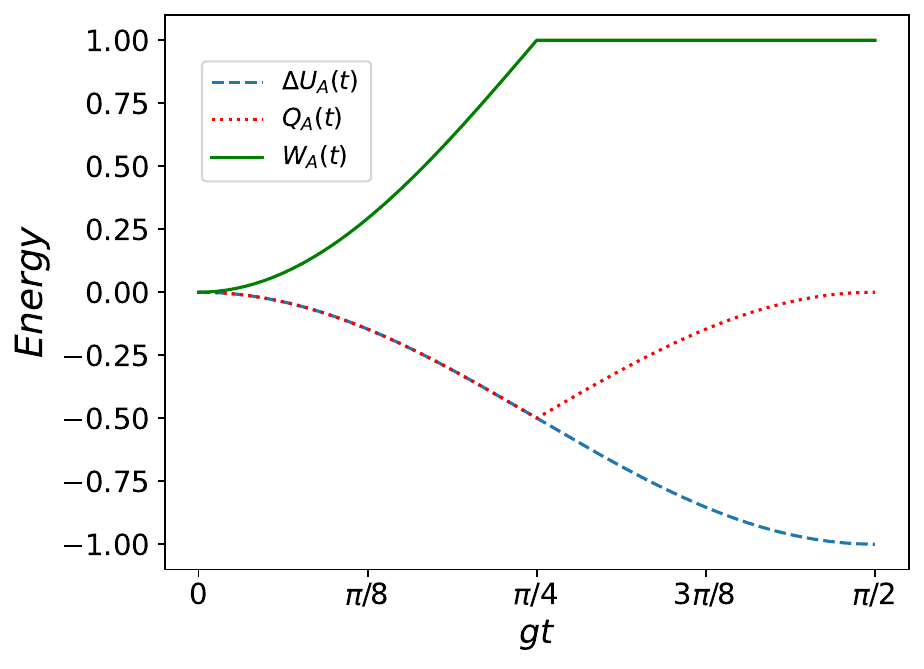}
        \caption{Qubit (System A).}
        \label{fig:sub1}
    \end{subfigure}
    
    \vspace{0.5cm} 

    \begin{subfigure}[b]{0.8\linewidth}
        \centering
        \includegraphics[width=\linewidth]{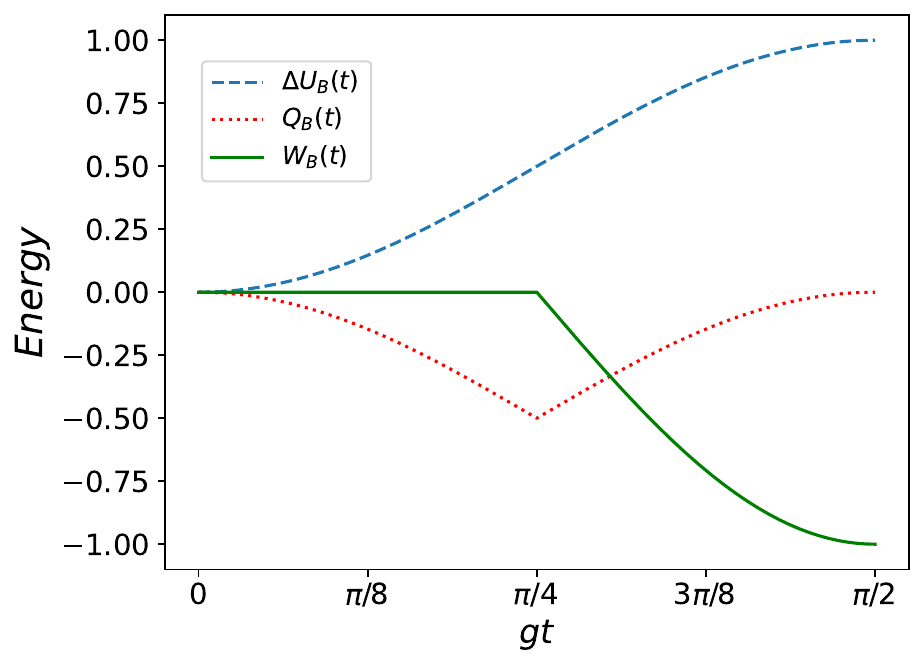}
        \caption{Qubit (System B).}
        \label{fig:sub2}
    \end{subfigure}

    \caption{Plot of the internal energy, heat and work for both qubits, as a function of time. The qubit A is initially in the excited state, $\ket{e}$ and qubit B is initially in the ground state, $\ket{g}$. Parameters: $\omega_A = \omega_B$ and $g = 0.01 \omega_A$ }
    \label{fig1}
\end{figure}

Similar to what we did in the qubit–QHO case, we will restrict our analysis to the time interval \( 0 \leq t \leq \tau = \pi / 2g \), because in this case as well the dynamics, and consequently the energy exchange, are periodic. First of all, the thermodynamic flows of qubit A in Fig.~\ref{fig:sub1} coincide with those of Fig.~\eqref{fig:sub31}. This is simply because the dynamics of qubit A is identical in both scenario.
On the other hand, we observe that the thermodynamic quantities associated with qubit B differ from those obtained for the QHO in Fig.~\ref{fig:sub32}. First, the magnitude of the heat absorbed or released by qubit B is greater than that of the QHO, and its heat exchange displays a sharp transition at \( t = \pi / 4g \). These differences stem from the distinct Hilbert space dimensions of system B in each scenario.

Regarding work production, for \( t \leq \pi / 4g \) no work is received by system B, since during this interval qubit B remains in a thermal state, unlike the QHO. For \( \pi / 4g \leq t \leq \tau=\pi/2g \), however, work is indeed performed on qubit B. This is because during this time interval, population inversion is generated, which illustrates that it requires work and that it is not a free resource.

\begin{figure}
    \centering
    \includegraphics[width=0.8\columnwidth]{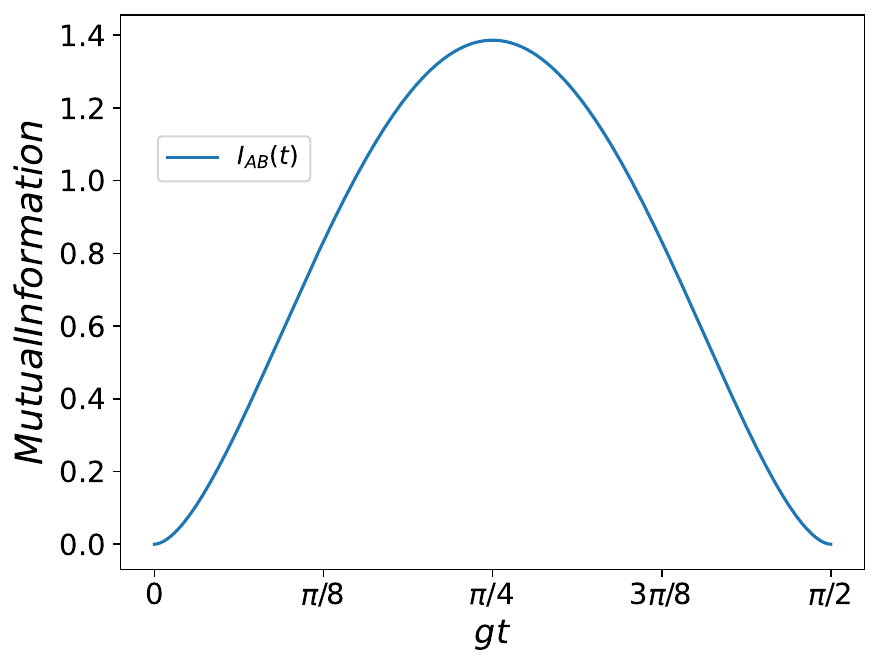}
    \caption{Plot of the mutual information between systems A and B as a function of time. Parameters: $\omega_A = \omega_B$ and $g = 0.01 \omega_A$.}
    \label{MI}
\end{figure}

Still in this example (Fig. \ref{fig1}), we can say that the work initially provided by A is converted into correlations between A and B (see Fig. \ref{MI}), which degrades the purity of A and B, resulting in incoming heat for both. In a second time, the heat provided by A and B is converted into work received by B, while the correlations between A and B are consumed. This work-information conversion creates an effect of delay between the moment in which A provides work and the moment in which B receives it.

As a second example, let us consider qubit A initially prepared in a superposition state, $\ket{\psi_A(0)} = \frac{1}{\sqrt{2}}(\ket{e} + \ket{g})$, while qubit B is still initially in the ground state.
\begin{figure}[h!]
    \centering
    \begin{subfigure}[b]{0.8\linewidth}
        \centering
        \includegraphics[width=\linewidth]{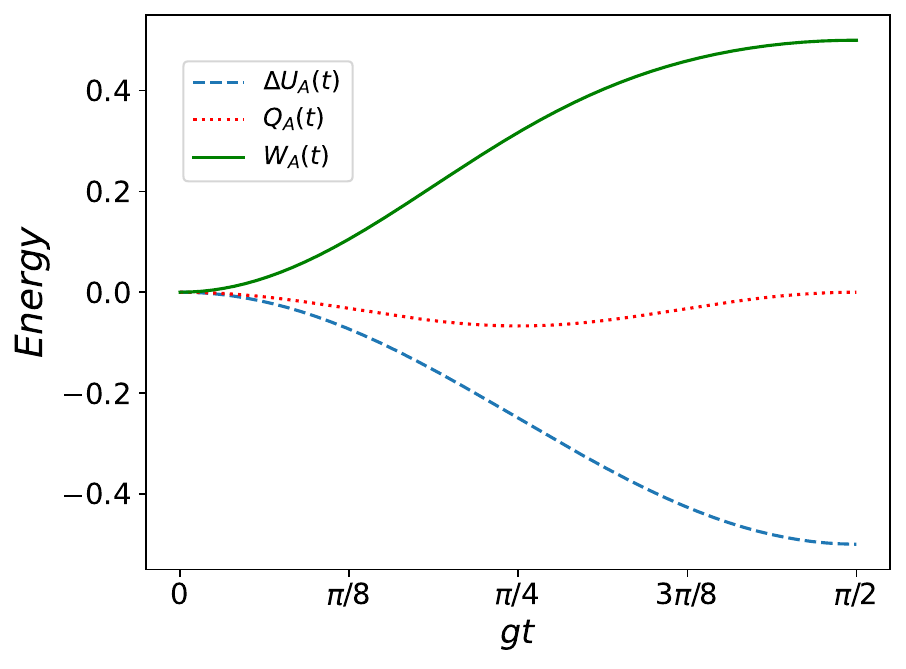}
        \caption{Qubit (System A).}
        \label{fig:sub21}
    \end{subfigure}
    
    \vspace{0.5cm} 

    \begin{subfigure}[b]{0.8\linewidth}
        \centering
        \includegraphics[width=\linewidth]{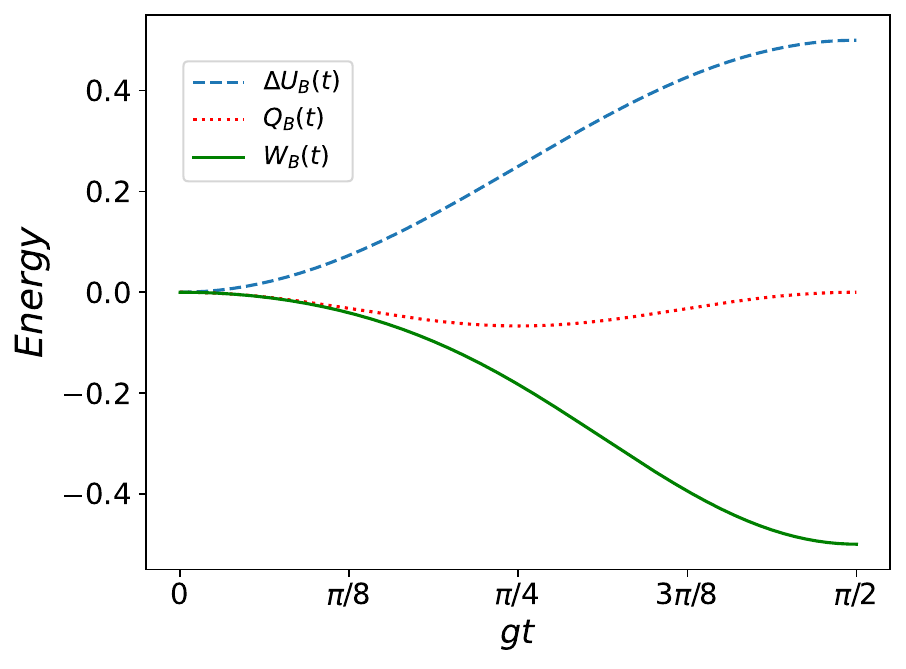}
        \caption{Qubit (System B).}
        \label{fig:sub22}
    \end{subfigure}

    \caption{Plot of the internal energy, heat and work for both qubits as a function of time. The qubit A is initially in a superposition state, $\ket{\psi_A(0)} = \frac{1}{\sqrt{2}}(\ket{e} + \ket{g})$ and qubit B is initially in the ground state, $\ket{g}$, again. Parameters: $\omega_A = \omega_B$ and $g = 0.01 \omega_A$. }
    \label{fig2}
\end{figure}
In Fig. \ref{fig2}, we observe that the thermodynamic quantities for both systems display the same behavior as those shown in Fig. \ref{fig4}, highlighting that the dimension of the Hilbert space plays no role here. This occurs because, in this situation, the dominant mechanism for work production stems from the coherence initially present in the qubit A. Therefore, the coherence generated within the subspace $ \{\ket{0}, \ket{1}\}$ of the QHO and  the subspace $ \{\ket{g}, \ket{e}\}$ of the qubits plays the key role to produce or generate work.

It is worth noting that, for a two-level system, the second term on the right side of Eq.~\eqref{ex} is always zero. It happens because for a two-level system any passive state is necessarily a thermal state. Therefore, for a qubit, the work done by or on the system is always equals to the ergotropy of the system.
In these qubit-qubit examples, we saw again the two unitary mechanisms of work production, the role of the Hilbert space dimension for the population-inversion mechanism (and its absence of role for the coherence-based mechanism), as well as the delay mechanism enabled by work-correlation conversion.

In the next example, we replace the quantum system B by a thermal bath of zero-temperature and we will analyze the thermodynamics of the spontaneous emission phenomenon.

\subsubsection{Spontaneous emission}

Now, let us consider the phenomenon of spontaneous emission. A qubit initially prepared in the excited state, $\ket{e}$, is brought into contact with a thermal reservoir at zero temperature and will asymptotically relax to the ground state, $\ket{g}$. In this scenario, the system only exchanges energy with the thermal bath, and and its dynamics can be described by a master equation in Lindblad form Ref.~\cite{open},
\begin{equation}
    \dot{\rho}_q(t) = - \frac{i}{\hbar}[H_q,\rho_q(t)] + \Gamma (\sigma_-\rho_q(t)\sigma_+ - \frac{1}{2}\{\sigma_+\sigma_-,\rho_q(t)\}),
\end{equation}
where $H_q = \frac{\hbar \omega}{2}\sigma_z$ is the qubit Hamiltonian, $\sigma_- = \ket{g}\bra{e}$, $\sigma_+ = \sigma_-^{\dagger}$, $\rho_S(t)$ is the density matrix of the qubit, and $\Gamma$ is the transition rate induced by the coupling with the thermal bath. Solving the dynamics of the system, we obtain 
\begin{align}
    P_e(t) &= e^{-\Gamma t} \\
    P_g(t) &= 1 - e^{-\Gamma t},
\end{align}
where $P_e(t)$ is the population of the excited state and $P_g(t)$ is the population of the ground state. It is worth noting that, in this example, the system remains diagonal at all times, $\rho_q(t) = P_e(t) \ket{e}\bra{e} + P_g(t) \ket{g}\bra{g}$. 

From the standard definitions of work and heat in quantum thermodynamics, Eqs. \eqref{wst} and \eqref{qst}, we see that there is no work production (i.e., $\dot{W}_q^{st} = 0$), since $H_q$ is time-independent. Consequently, there is only heat exchange between the system and the thermal bath, and it is given by
\begin{equation}\label{qspst}
    Q_q^{st}(t) = \omega\left[P_e(t) - 1\right].
\end{equation}
This indicates that, according to the standard framework, the system always provides heat to the thermal bath. One reaches the same conclusion using the framework \cite{esp}.

By contrast, using the autonomous framework \cite{1}, we see that for $0 \leq t \leq \tau$, where $\tau$ is the instant of time at which $P_e(\tau) = P_g(\tau) = 1/2$, we have $P_e(t) \geq P_g(t)$, implying, according to Eq. \eqref{ent}, $\rho_q^{th} =  P_e(t) \ket{g}\bra{g} + P_g(t)\ket{e}\bra{e}$. In this scenario, we obtain from Eqs. \eqref{heat} and \eqref{exergy3}
\begin{align}\label{qsp}
    Q_q(t) &= -\omega P_g(t) = -\omega [1 - P_e(t)] \\
    W_q(t) &= 2 \omega[1 - P_e(t)]\label{wsp},
\end{align}
and $\Delta U_q(t) = \omega[P_e(t) - 1]$. Following our signal convention, the system is absorbing heat from the bath and providing work to it.

\begin{figure}
 \begin{subfigure}[b]{0.9\linewidth}
        \centering
        \includegraphics[width=\linewidth]{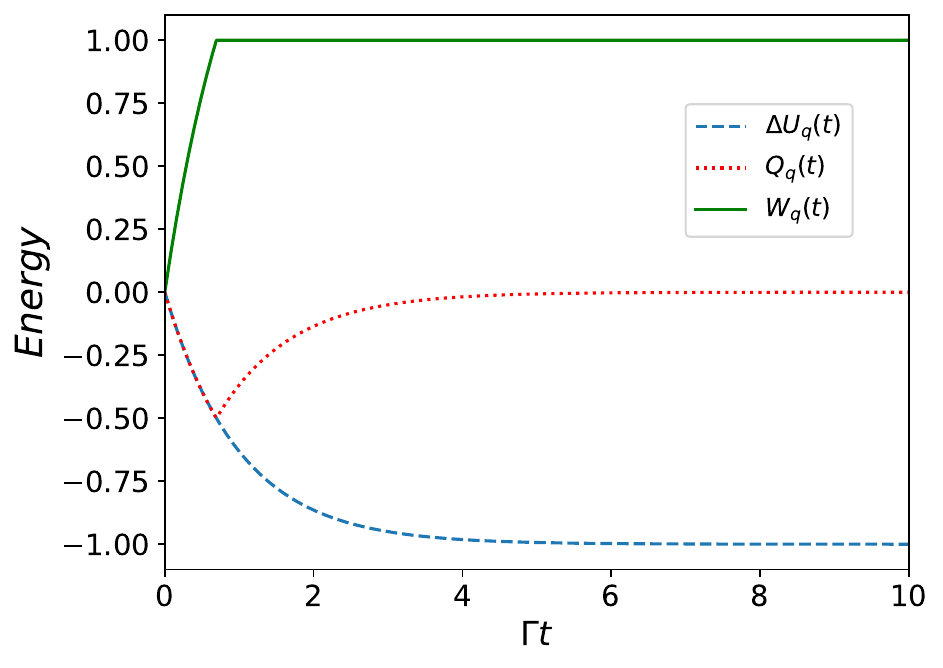}
        \label{fig:sub61}
    \end{subfigure}
    \caption{Plot of the thermodynamic quantities as a function of time for qubit spontaneous emission. The qubit is initially in the excited state $\ket{e}$. Parameters: $\Gamma = 0.01 \omega$.}
    \label{fig6}
\end{figure}

Then, for $t > \tau$, we have $P_e(t) \leq P_g(t)$, and thus from Eq. \eqref{ent}, $\rho_q^{th} =  P_g(t) \ket{g}\bra{g} + P_e(t)\ket{e}\bra{e}$. In this case, there is no work production and the heat exchange between the system and the thermal bath is given by 
\begin{equation}
    Q_q(t) = \omega\left[P_e(t) - 1\right],
\end{equation}
and it coincides with the standard heat definition Eq. \eqref{qspst}. $W_q$ and $Q_q$ are plotted in Fig. \ref{fig6}.

These results highlight a clear distinction in the thermodynamic analysis between the standard framework and the autonomous framework described in Sec.~\ref{TF}. In the former, the system exchanges only heat with the thermal bath at all times, whereas in the latter, for $0 \leq t \leq \tau$, the system performs work through the population-inversion mechanism.

It is also worth noting that, during the interval $0 \leq t \leq \tau$, the system absorbs heat from the bath due to the system-bath correlation build-up, rather than releasing it as in the standard formalism. Consequently, what is interpreted as pure heat exchange in the standard framework corresponds to a combination of work performed by the system on the thermal bath and heat absorbed by the system.

Finally, note that the thermodynamic behavior showed in Fig. \ref{fig6} is similar to that shown in Fig. \ref{fig1}. The difference lies in the fact that in Fig. \ref{fig1} the thermodynamic quantities are periodic, while here they reach asymptotic value, due to the thermalization of the qubit with the thermal bath.

Next, in order to have access to the thermodynamics of the bath, as well as to treat strong system-bath coupling, we use the Reaction Coordinate (RC) method \cite{Garg1985,Strasberg2018, Perarnau2018, Iles2014, Iles2018, Latune2022}. In this approach, the degrees of freedom of the bath that are most strongly coupled to the qubit are mapped onto an explicit quantized mode—the reaction coordinate—which is treated as part of an enlarged system (See appendix \ref{AppA3} for more details).

Accordingly, we now consider a qubit interacting with a quantized mode, which in turn interacts with a zero-temperature Markovian bath. In this scenario, the total dynamics is ruled by a master equation~\cite{Iles2014,Iles2018, SPME2, Latune2022}
\begin{equation}\label{RCME}
    \dot{\rho}(t) = -\frac{i}{\hbar}\hskip 0.02cm[H_{Q,RC},\rho(t)] + \frac{\kappa}{2}\hskip 0.02cm(2 a\rho(t)a^{\dagger} - \{a^{\dagger}a, \rho(t)\}),
\end{equation}
where $H_{Q,RC} = \frac{\hbar\omega}{2}\sigma_z + \hbar\omega_{RC} a^{\dagger}a + \hbar\lambda\hskip 0.02cm(\sigma_{+}a + \sigma_{-}a^{\dagger})$, $\{A, B\} = AB + BA$ and $\rho(t)$ is the joint state of the system. In the following, we assume that the qubit and the Reaction Coordinate are resonant,
$\omega = \omega_{\mathrm{RC}}$.

Then, to investigate the thermodynamics of the bath in the spontaneous emission process, we employ the reaction coordinate (RC) mapping, in which the RC represents the accessible part of the bath and allows us to extract thermodynamic information about it. In all cases considered in the following, the RC is initially prepared in the vacuum state $\ket{0}$, while the qubit is initially in the excited state $\ket{e}$.
\begin{figure}[h!]
    \centering
    \begin{subfigure}[b]{0.8\linewidth}
        \centering
        \includegraphics[width=\linewidth]{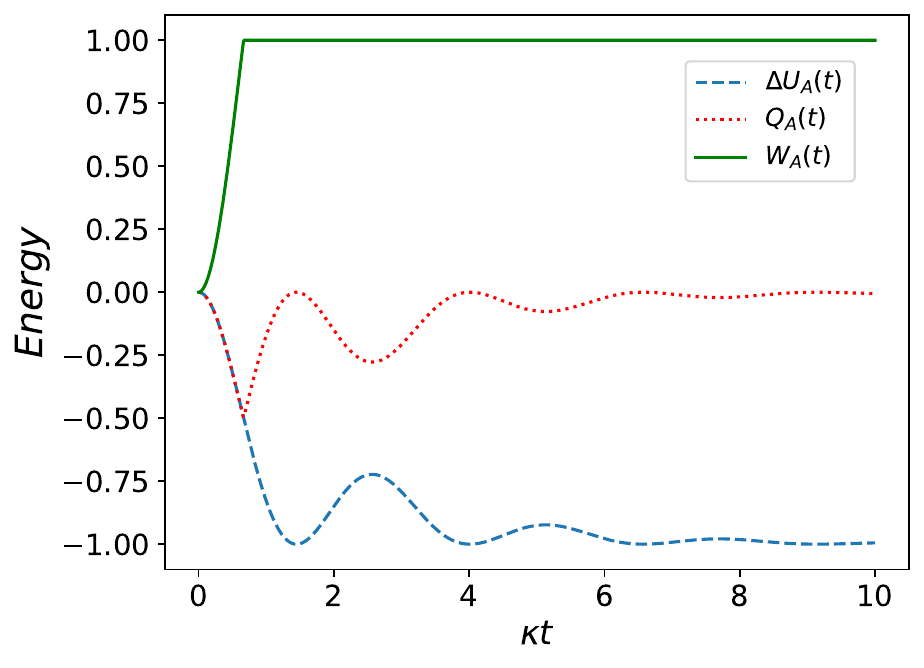}
        \caption{Qubit (System A).}
        \label{fig:subRC1}
    \end{subfigure}
    
    \vspace{0.5cm} 

    \begin{subfigure}[b]{0.8\linewidth}
        \centering
        \includegraphics[width=\linewidth]{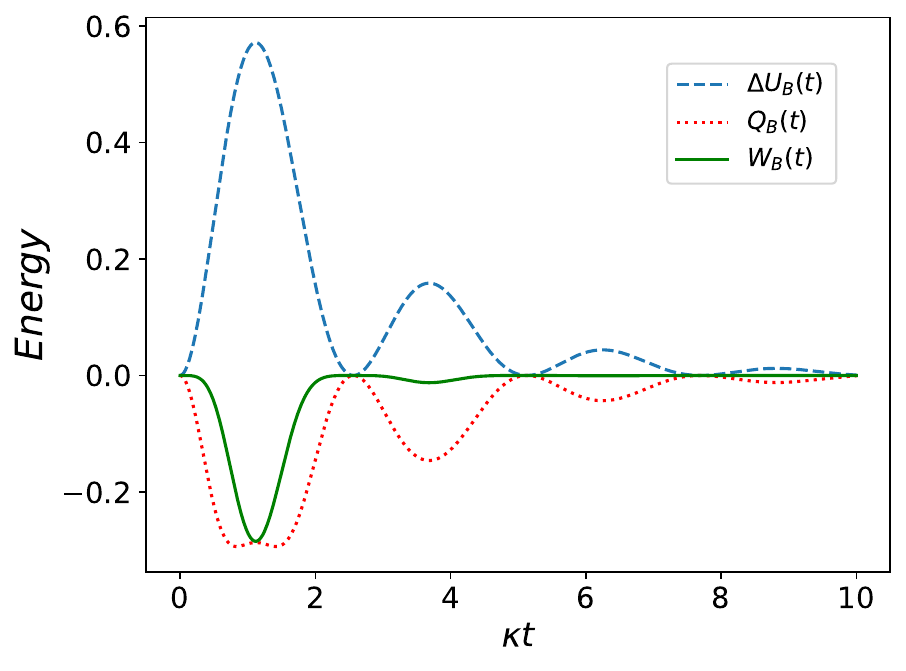}
        \caption{Reaction coordinate (System B).}
        \label{fig:subRC2}
    \end{subfigure}

    \caption{ Plot of the thermodynamic quantities of the qubit and the reaction coordinate as a function of time. Parameters: $\omega = \omega_{RC}$, $\kappa = 0.8\hskip 0.03cm \lambda$ and $\lambda = 0.01 \omega$. }
    \label{figRC}
\end{figure}

In Fig.~\ref{figRC}, we present the thermodynamic quantities of the qubit and the RC. The qubit is initially prepared in the excited state, while the quantized mode is initially in the vacuum state $\ket{0}$, corresponding to equilibrium with the reminder of the thermal bath (also called residual bath). We focus on an intermediate-coupling strength between the reaction coordinate and the residual bath, with $\kappa = 0.8\hskip 0.03cm \lambda$.

In Fig.~\ref{fig:subRC1}, we observe that the work supplied by the qubit is the same as the one in Fig. \ref{fig6}. However, due to the intermediate coupling strength regime, the provided heat flow as well as the energy of the qubit oscillate before being damped out, as a typical sign of non-vanishing coupling and non-Markovianity \cite{Iles2014, Iles2018, Latune2023}. 

In Fig.~\ref{fig:subRC2}, we observe that the reaction coordinate indeed receives part of the work supplied by the qubit, contrasting again with the standard thermodynamic framework which reduces all energy exchanges to heat. Still, there are some heat exchanges with both the qubit and the residual bath, but represent only a fraction of the exchanged energy. At long time, both work and heat received by the Reaction Coordinate are damped in the remainder of the bath.

In conclusion, this example shows that the bath indeed receives part of the work provided by the qubit, but this energy is rapidly dissipated through the modes of the full bath. Naturally, changing the coupling strength between the reaction coordinate and the bath affects both the amount of work transferred to the bath and the timescale over which it is dissipated (see Appendix~\ref{AppA3} for additional examples).

\section{Conclusion and discussion}

In this work, we applied a thermodynamic formalism \cite{1} designed for autonomous quantum systems to study the thermodynamics of interacting quantum systems in both semiclassical and quantum regimes.

In the semiclassical regime, we showed that when two systems interact but one of them behaves classically, energy exchange occurs purely as work production. Comparing the framework from \cite{1} with standard definitions \cite{Spohn1978, Alicki1979} and other recent approaches \cite{maffei, dann}, we found that the formalism \cite{1} identifies the local work contribution while accounting for the interaction energy separately. This distinction is particularly interesting for systems in which the interaction energy plays a non-negligible role—for instance, when the cost of switching the interaction on and off cannot be neglected.
We illustrate this point with the thermodynamic analysis of a two-level system driven by a quantum harmonic oscillator initially in a large coherent state, reproducing the experimental situation of a qubit driven by a laser. We show that the energy exchanged between the qubit and the laser (harmonic oscillator) is indeed work, until the semi-classical approximation starts to breakdown with the emergence of correlations implying a slowly increasing incoming heat flow.

In the quantum regime, our results point to many mechanisms of work production. We identified two unitary mechanisms, one based on population inversion and the second based on quantum coherences. These unitary mechanisms have been associated with work cost in many papers \cite{Misra2016, Kallush2019, Marvian2020, Hadipour2024} including as main mechanism of ergotropy production \cite{Ergotropy, Niu2024, Lutz2024}, but, curiously, are not detected by other thermodynamic frameworks, including by the standard ones. On top of that, our results point to an extra work production mechanism, which is a non-unitary mechanism and occurs when systems of different dimensions interact. This indicates that Hilbert space dimensionality can serve as an intrinsic resource for work generation in autonomous quantum systems.
Finally, we also shows that the two unitary mechanisms of work production can also occurs with an infinite thermal bath, even though the work received by the bath is at long times eventually dissipated in all bath modes.

We believe that our results show the versatility of the autonomous framework and contribute to the understanding of the work and heat exchange mechanisms in the quantum domain, and particularly in autonomous settings. This is essential to harness the different forms of energy at the quantum level, aiming at keeping under control the energy consumption of quantum devices and quantum technologies.

\acknowledgements
The authors acknowledge funding from the French National Research Agency (ANR) under grant ANR-23-CPJ1-0030-01.

\appendix
\section{Exergy}
\label{AppA1}
In the main text, we defined $\Xi_j(t) = \operatorname{Tr}\{[\rho_j(t) - \rho_j^{th}(t)]H_j\}$. By adding the passive energy $U_P(t) = \operatorname{Tr}[\pi_j(t)H_j]$, associated with the passive state $\pi_j(t) = \sum_i r_i(t)\ket{E_i}\bra{E_i}$, to this, we obtain
\begin{equation}\label{A1}
    \Xi_j(t) = \operatorname{Tr}\{[\rho_j(t) - \pi_j(t)]H_j\} + \operatorname{Tr}\{[\pi_j(t) - \rho_j^{th}(t)]H_j\}.
\end{equation}
The first term in the right hand of the equation above is the ergotropy, then, we can rewrite the above equation as
\begin{equation} \label{A2}
\Xi_j(t) = \mathcal{E}_j(t) + \operatorname{Tr}{[\pi(t) - \rho_j^{th}(t)]H_j}.
\end{equation}

Now, Substituting $H_j = -\frac{1}{\beta_j(t)}\ln[\rho_j^{th}(t) Z_j(t)]$ into Eq.~\eqref{A2} and adding $\frac{S[\pi(t)] - S[\pi(t)]}{\beta_j(t)}$ to its right-hand side, we obtain, after some algebraic manipulation,
\begin{equation}\label{A3}
    \Xi_j(t) = \mathcal{E}_j(t) + \frac{1}{\beta_j(t)} D[\pi_j(t)||\rho_j^{th}(t)],
\end{equation}
where the second term on the right hand side is the relative quantum entropy. We call this quantity as Exergy~\cite{exergy1, exergy2}, and as both the ergotropy, $\mathcal{E}_j(t)$, and the relative quantum entropy $D[\pi_j(t)||\rho_j^{th}(t)]$ are always positve, the exergy is always positive, $\Xi_j(t) \geq 0$

Eq. \eqref{A3} shows that the exergy, $\Xi_j(t)$, is always greater than or equal to the ergotropy, $\mathcal{E}_j(t)$,and reveals two distinct mechanisms for work generation: a unitary mechanism associated with the production or extraction of ergotropy, and a nonunitary mechanism associated with the thermal background (characterized by the inverse effective temperature $\beta_j(t))$) and with the distance between the passive state, $\pi_j(t)$, and the thermal state, $\rho_j^{th}(t)$.

\section{Qubit-QHO extra comments} \label{AppA2}
\begin{figure}[!t]
    \centering
    \begin{subfigure}[b]{0.8\linewidth}
        \centering
        \includegraphics[width=\linewidth]{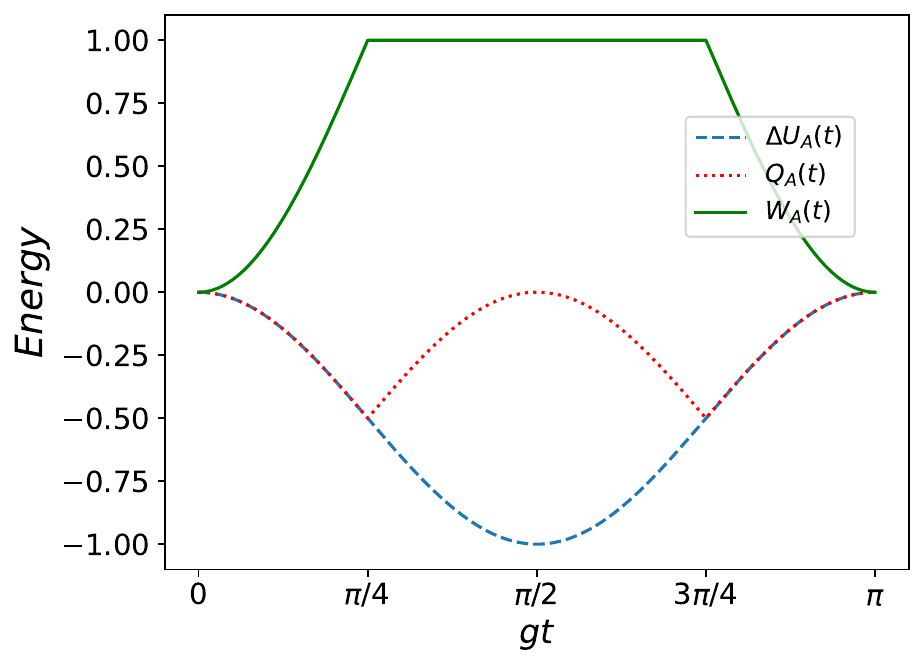}
        \caption{Qubit (System A).}
        \label{fig:subAppB11}
    \end{subfigure}
    
    \vspace{0.5cm} 

    \begin{subfigure}[b]{0.8\linewidth}
        \centering
        \includegraphics[width=\linewidth]{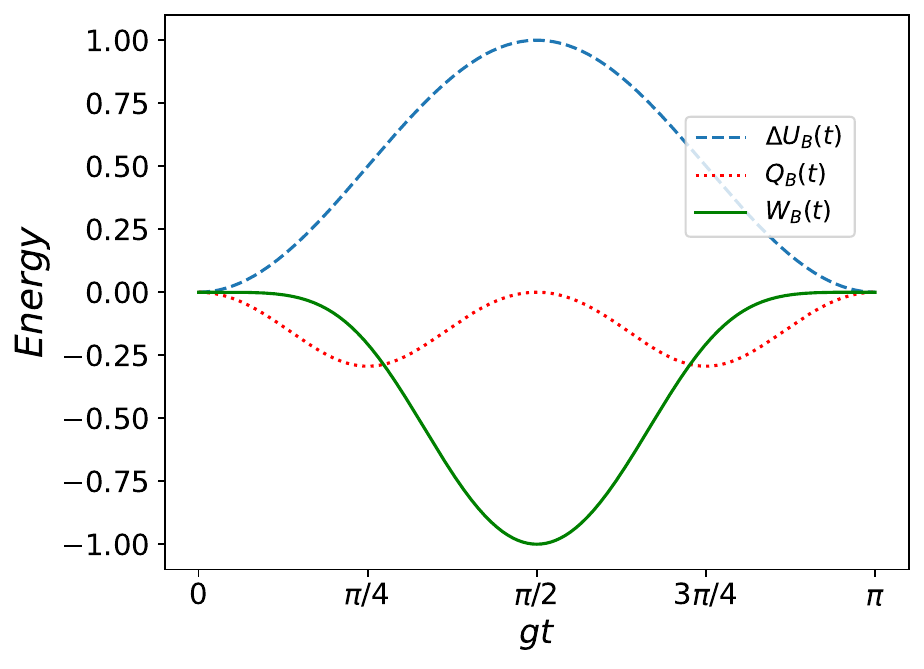}
        \caption{QHO (System B).}
        \label{fig:subAppB12}
    \end{subfigure}

    \caption{Here, we plotted the the thermodynamical quantities for both the qubit (A) and the QHO (B), as a function of time . The qubit is initially in the excited state, $\ket{e}$, and the QHO is initially in the vacuum state, $\ket{0}$. Parameters: $\omega_A = \omega_B$ and $g = 0.01 \omega_A$.}
    \label{figAppB1}
\end{figure}
In the main text, we analyzed the thermodynamics of the qubit--QHO and qubit--qubit systems in the time interval $0 \leq t \leq \pi/2g$, except when the QHO is initially prepared in the coherent state $\ket{\alpha = 30}$. This restriction was justified by the periodic nature of the dynamics and, consequently, of the thermodynamic behavior. However, the interval $0 \leq t \leq \pi/2g$ corresponds to only half of the full period, $0 \leq t \leq \pi/g$. For the complementary interval $\pi/2g \leq t \leq \pi/g$, the dynamics correspond to a reflection of the evolution in $0 \leq t \leq \pi/2g$ about the vertical line at $t = \pi/2g$. As a consequence, if a system provides (receives) work or heat in the interval $0 \leq t \leq \pi/2g$, it will receive (provide) the same amount of work or heat in the interval $\pi/2g \leq t \leq \pi/g$. This behavior arises because, at $t = \pi/2g$, the systems exchange their states and their role and subsequently evolve back to their respective initial states in the interval $\pi/2g \leq t \leq \pi/g$.

In Fig.~\ref{figAppB1}, which corresponds to the full-period plot of Fig.~\ref{fig3}, we observe that, for $\pi/2g \leq t \leq \pi/g$, the thermodynamic quantities behave as if the systems were following a reversed path relative to the interval $0 \leq t \leq \pi/2g$, evolving from $t = \pi/2g$ back to $t = 0$. For example, in the qubit case, from $t = \pi/2g$ to $t = 3\pi/4g$ there is no work production, in analogy with the interval $t = \pi/4g$ to $t = \pi/2g$, while from $t = 3\pi/4g$ to $t = \pi/g$ the qubit receives the same amount of work that it provided at early times, from $t = 0$ to $t = \pi/4g$.

For the heat exchange, we observe a symmetry between the corresponding time intervals: in both cases, heat is initially absorbed by the qubit and subsequently released. This occurs because, in both intervals, the qubit starts in a pure state, evolves to a maximally mixed state—requiring heat absorption over a time interval $\Delta t = \pi/4g$—and finally evolves to a pure state again, leading to heat release.

\begin{figure}[!t]
    \centering
    \begin{subfigure}[b]{0.7\columnwidth}
        \centering
        \includegraphics[width=\linewidth]{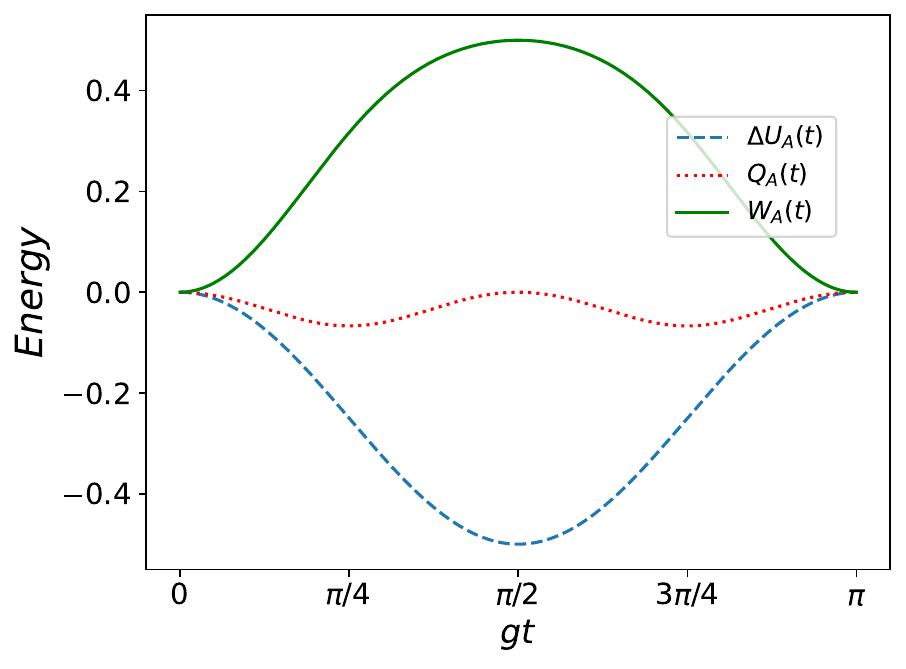}
        \caption{Qubit (system A).}
        \label{fig:subappB21}
    \end{subfigure}
    
    \vspace{0.5cm} 

    \begin{subfigure}[b]{0.7\linewidth}
        \centering
        \includegraphics[width=\linewidth]{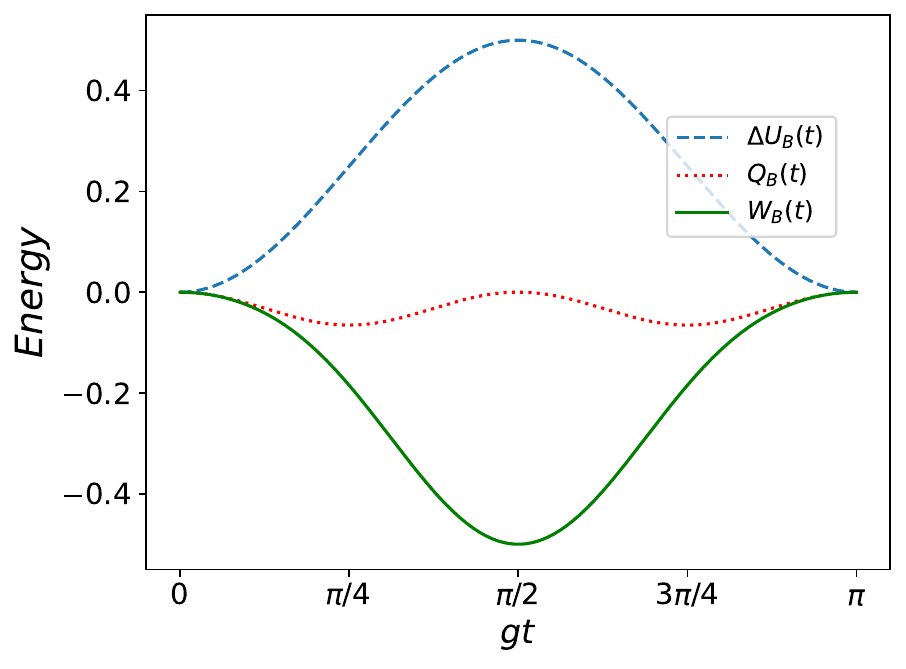}
        \caption{Qubit B.}
        \label{fig:subappB22}
    \end{subfigure}

    \caption{Here, we plotted the internal energy, heat and work for both systems as a function of time. The qubit (A) is initially in a superposition state, $\ket{\psi_A(0)} = \frac{1}{\sqrt{2}}(\ket{e} + \ket{g})$ and the QHO (B) is initially in the vacuum state, $\ket{0}$. Parameters: $\omega_A = \omega_B$ and $g = 0.01 \omega_A$. }
    \label{figAppB2}
\end{figure}

In Fig.~\ref{figAppB2}, which corresponds to the full-period plot of Fig.~\ref{fig4}, we observe that, for $\pi/2g \leq t \leq \pi/g$, the thermodynamic quantities form the mirror image of their behavior in the interval $0 \leq t \leq \pi/2g$, i.e., it is a reflection about the vertical line at $t=\pi/2g$. For the qubit (A), for instance, work is provided in the interval $0 \leq t \leq \pi/2g$, whereas in the interval $\pi/2g \leq t \leq \pi/g$ it receives the same amount of work that was provided in the first half-period. For the QHO, we observe the opposite behavior: it receives work in the interval $0 \leq t \leq \pi/2g$ and provides work in the interval $\pi/2g \leq t \leq \pi/g$. The heat exchange for both systems follows the same behavior discussed in the previous paragraph for Fig.~\ref{figAppB1}.

\section{Reaction coordinate: additional information}\label{AppA3}

Consider a qubit (Q) interacting with a zero-temperture bath (E), according to the following Hamiltonian (in the following $\hbar = 1$)
\begin{equation}
    H_{QE} = \frac{\omega}{2}\sigma_z + \sum_{k} \omega_k a_k^{\dagger} a_k + \sum_{k} \Omega_k\hskip 0.02cm(\sigma_{+}a_k + \sigma_{-}a^{\dagger}_k).
\end{equation}
We can rewrite the above Hamiltonian as
\begin{equation}
    H_{QE} = \frac{\omega}{2}\sigma_z + \sigma_+E+\sigma_-E^\dag+ \sum_k \omega_ka_k^\dag a_k
\end{equation}
with $E:=\sum_k\Omega_ka_k$. For simplicity we are assuming that $\Omega_k$ is Real. Introducing the interaction picture operator $E^\dag(t) := e^{i H_Et/\hbar} E^\dag e^{-i H_Et/\hbar}$, where $H_E = \sum_k \omega_k a_k^\dagger a_k$, the bath correlation function is
\begin{equation}
c(t) :=  {\rm Tr}[\rho_B E^\dag(t) E] = \sum_k \Omega_k^2 e^{i\omega_k t} n_k
\end{equation}
with $n_k := {\rm Tr}[\rho_Bb_k^\dag b_k]$. The correlation function is usually computed going to the continuum of mode by introducing the spectral density $D(\omega)$
\begin{equation}
\sum_k  \Omega_k^2  \rightarrow \int_0^\infty d\omega \frac{\Lambda^2}{2\pi} D(\omega)
\end{equation}
such that $\int_0^\infty d\omega D(\omega) = 2\pi$ and therefore $\Lambda$ represents the coupling strength with the bath.

Taking a Lorentzian spectral density (sometimes also called under-damped), 
\begin{equation}
D(\omega) = \omega\frac{2}{\pi}\frac{\gamma\lambda^2\omega_c^2}{(\omega^2-\omega_c^2)^2+(\omega \gamma\omega_c)^2}
\end{equation}
where $\omega_c$ corresponds to the peak of the Lorentzian while $\gamma$ describes its width and $\lambda$ the bath coupling strength, the above problem can be exactly mapped onto \cite{Garg1985,Strasberg2018, Perarnau2018, Iles2014, Iles2018, Latune2022} our system $Q$ increased by a fictitious bosonic mode called reaction coordinate RC, itself ineracting with a residual Markovian bath at zero temperature. In the limit of weak coupling with the residual bath, which corresponds to the situation of a narrow spectral density, the dynamics of $Q$ plus RC is given by

\begin{equation}
    \dot{\rho}(t) = -i\hskip 0.02cm[H_{Q,RC},\rho(t)] + \frac{\kappa}{2}\hskip 0.02cm(2 a\rho(t)a^{\dagger} - \{a^{\dagger}a, \rho(t)\}),
\end{equation}
where  $\{A, B\} = AB + BA$, $H_{Q,RC} = \frac{ \omega}{2}\sigma_z +  \omega_{RC} a^{\dagger}a + \lambda\hskip 0.02cm(\sigma_{+}a + \sigma_{-}a^{\dagger})$, with $a$, $a^\dag$ being respectively the annihilation and creation operator of the Reaction Coordinate, and $\omega_{RC} = \omega_c$. Note that the damping rate $\kappa$ is determined by $\gamma$, the width of the bath spectral density. 
In the following we will assume that the qubit and the RC are resonant $(\omega = \omega_c)$.

To investigate the thermodynamics of the bath in the spontaneous emission process, we employ the reaction coordinate (RC) mapping, in which the RC represents the accessible part of the bath and allows us to extract thermodynamic information about it. In all cases considered in the following, the RC is initially prepared in the vacuum state $\ket{0}$, while the qubit is initially in the excited state $\ket{e}$.

\begin{figure}[h!]
    \centering
    \begin{subfigure}[b]{0.8\linewidth}
        \centering
        \includegraphics[width=\linewidth]{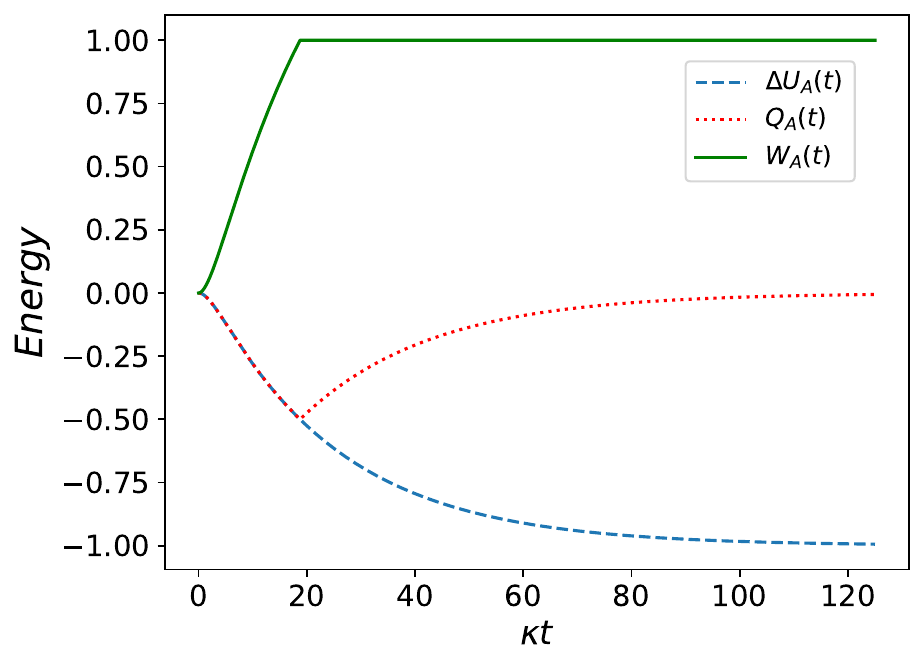}
        \caption{Qubit (System A).}
        \label{fig:appRC1}
    \end{subfigure}
    
    \vspace{0.5cm} 

    \begin{subfigure}[b]{0.8\linewidth}
        \centering
        \includegraphics[width=\linewidth]{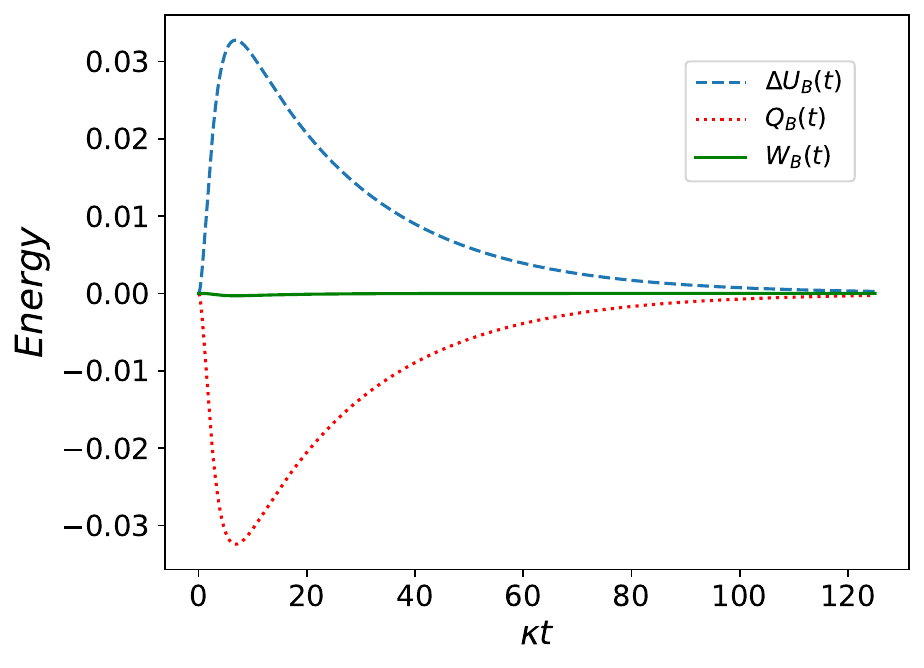}
        \caption{Reaction coordinate (System B).}
        \label{fig:appRC2}
    \end{subfigure}

    \caption{thermodynamic quantities of the qubit and the quantized mode (RC) as a function of time. Parameters: $\omega = \omega_{RC}$, $\kappa = 10\hskip 0.03cm \lambda$ and $\lambda = 0.01 \omega$. }
    \label{figRCapp}
\end{figure}

In Fig.~\ref{fig:appRC1}, we present the thermodynamic quantities of the qubit and the reaction coordinate in the strong-coupling regime between the RC and the residual bath. As shown in Fig.~\ref{fig:appRC1}, the thermodynamic behavior of the qubit is very similar to that obtained in Fig.~\ref{fig6}, as it should be. For the RC, we observe that it receives a very small amount of work at early times, over a short time interval, and exchanges a small amount of heat with both the qubit and the bath. At long times, as expected, all thermodynamic quantities are dissipated into the residual bath. The amount of energy exchanged by the RC in this case is smaller than in the intermediate-coupling case shown in Fig.~\ref{figRC}, because the damping of energy exchange increases with increasing $\kappa$. Note also that, for $\kappa \gg \Omega$, the dynamics effectively correspond to a qubit directly coupled to a zero-temperature bath, such that the cavity can be eliminated from the description.

\begin{figure}
    \centering
    \begin{subfigure}[b]{0.8\linewidth}
        \centering
        \includegraphics[width=\linewidth]{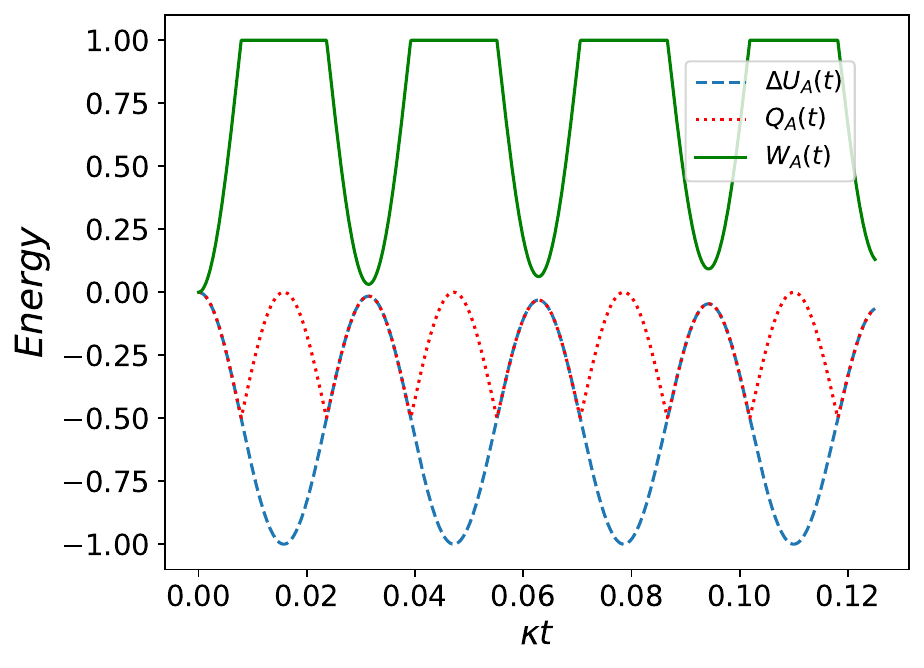}
        \caption{Qubit (System A).}
        \label{fig:appRC3}
    \end{subfigure}
    
    \vspace{0.5cm} 

    \begin{subfigure}[b]{0.8\linewidth}
        \centering
        \includegraphics[width=\linewidth]{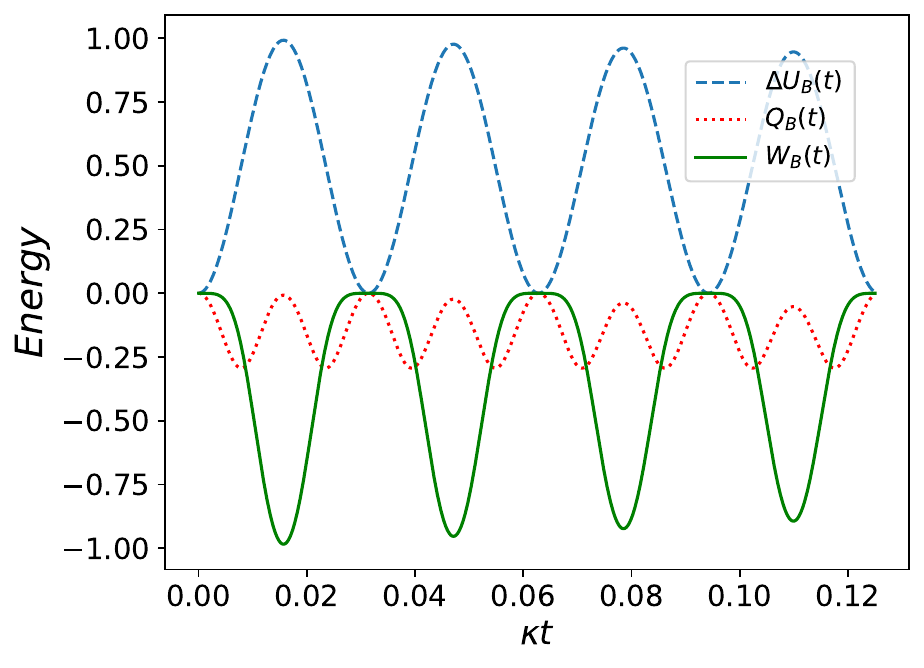}
        \caption{Reaction coordinate (System B).}
        \label{fig:appRC4}
    \end{subfigure}

    \caption{thermodynamic quantities of the qubit and the quantized mode (RC) as a function of time. Parameters: $\omega = \omega_{RC}$, $\kappa = 0.01\hskip 0.03cm \lambda$ and $\lambda = 0.01 \omega$. }
    \label{figRCapp2}
\end{figure}

We now consider the weak-coupling limit between the reaction coordinate and the bath. In this regime, Fig.~\ref{figRCapp2} shows that the thermodynamic quantities become periodic. However, for the qubit (Fig.~\ref{fig:appRC3}), the minimum value of the work at the end of each period increases over time. A similar behavior is observed for the internal energy, but in terms of its maximum value. For the RC (Fig.~\ref{fig:appRC4}), we observe a related trend: the minimum value of the work increases from period to period, while the maximum value of the internal energy decreases over successive periods. This behavior arises from the influence of the bath, which, although weakly coupled to the RC, still affects the energy exchange of the qubit and the RC. At very long time, such oscillating behavior tends to fade out, as in previous situations. Nevertheless, for sufficiently small values of the dissipation rate $\kappa$, the dynamics approach those of a system effectively decoupled from the bath, and the qubit exhibits perfectly periodic thermodynamic behavior.

\end{document}